\documentclass[aps,prx,twocolumn,superscriptaddress,showpacs,floatfix]{revtex4-2}

\usepackage{amsmath,amssymb,amsfonts,float,graphics,epsfig,epstopdf,color,verbatim,tabularx,bm,multirow,appendix}
\usepackage[utf8]{inputenc}
\usepackage[T1]{fontenc}
\usepackage{xcolor}
\usepackage{dsfont}
\usepackage{textcomp}
\usepackage{yfonts}
\usepackage{footnote}
\usepackage{bm}
\usepackage{bbm}
\usepackage{subfigure}
\usepackage{tabularx}
\usepackage{mathrsfs}
\usepackage{graphicx}
\usepackage{verbatim}
\usepackage[colorlinks=true, citecolor=blue, linkcolor=blue, urlcolor=blue]{hyperref}
\usepackage{multirow}
\usepackage{braket}
\usepackage[normalem]{ulem}

\usepackage{tkz-euclide}
\tkzSetUpPoint[size=4,color=black]
\tikzset{new/.style={color=black,line width=0.5pt}}

\usetikzlibrary{calc}
\usetikzlibrary{shapes.multipart}
\usepackage{hyperref}
\hypersetup{
    colorlinks=true,            
    linkcolor=blue,             
    filecolor=magenta,          
    urlcolor=cyan,              
    pdftitle={Overleaf Example},
    pdfpagemode=FullScreen,
    }
\usepackage{booktabs}
    
\graphicspath{ {images/} }

\DeclareMathOperator{\Tr}{Tr}

\newcommand{\comments}[1]{}

\newcommand{\av}[1]{\langle#1\rangle}

\newcommand{\eps}{\mathcal E}

\newcommand{\SvN}{S^{\mathrm{vN}}}

\newcommand{\stkout}[1]{\ifmmode\text{\sout{\ensuremath{#1}}}\else\sout{#1}\fi}


\begin{document}

\title{Multiparty Entanglement Microscopy of Quantum Ising models in 1d, 2d and 3d}

\author{Liuke Lyu}
\altaffiliation{The authors contributed equally to this work.}
\affiliation{D\'epartement de Physique, Universit\'e de Montr\'eal, Montr\'eal, QC H3C 3J7, Canada}

\author{Menghan Song}
\altaffiliation{The authors contributed equally to this work.}
\affiliation{Department of Physics and HK Institute of Quantum Science \& Technology, The University of Hong Kong, Pokfulam Road, Hong Kong SAR, China}

\author{Ting-Tung Wang}
\affiliation{Department of Physics and HK Institute of Quantum Science \& Technology, The University of Hong Kong, Pokfulam Road, Hong Kong SAR, China}

\author{Zi Yang Meng}
\email{zymeng@hku.hk}
\affiliation{Department of Physics and HK Institute of Quantum Science \& Technology, The University of Hong Kong, Pokfulam Road, Hong Kong SAR, China}

\author{William Witczak-Krempa}
\email{w.witczak-krempa@umontreal.ca}
\affiliation{D\'epartement de Physique, Universit\'e de Montr\'eal, Montr\'eal, QC H3C 3J7, Canada}
\affiliation{
 Institut Courtois, Universit\'e de Montr\'eal, Montr\'eal (Qu\'ebec), H2V 0B3, Canada
}
\affiliation{
 Centre de Recherches Math\'ematiques, Universit\'e de Montr\'eal, Montr\'eal, QC, Canada, HC3 3J7
}

\begin{abstract}
Entanglement microscopy reveals the true quantum correlations among the microscopic building blocks of many-body systems~\cite{wangEntanglement2024}.
Using this approach, we study the multipartite entanglement of the 
quantum Ising model in 1d, 2d, and 3d. We first obtain the full reduced density matrix (tomography) of subregions that have at most 4 sites via quantum Monte Carlo, exact diagonalization, and the exact solution in 1d. We then analyze both bipartite and genuine multipartite entanglement (GME) among the sites in the subregion. To do so, we use a variety of measures including the negativity, as well as a true measure of GME: the genuinely multipartite concurrence (or GME concurrence), and its computationally cheaper lower bound, $I_2$. We provide a complete proof that $I_2$ bounds the GME concurrence, and show how the symmetries of the state simplify its evaluation.
For adjacent sites, we find 3- and 4-spin GME  present across large portions of the phase diagram, reaching maximum near the quantum critical point. In 1d, we identify the singular scaling of the derivative $dI_2/dh$ approaching the critical point. 
We observe a sharp decrease of GME with increasing dimensionality, consistent with the monogamous nature of entanglement. Furthermore, we find that GME vanishes for subregions consisting of non-adjacent sites in both 2d and 3d, offering a stark illustration of the short-ranged nature of entanglement in equilibrium quantum matter~\cite{foe}. Finally, we analyze the most collective form of entanglement by evaluating the GME concurrence among all spins in the lattice, which can be obtained from a simple observable: the single-site transverse magnetization. 
This global concurrence is larger in 1d compared to 2d/3d, but it is relatively less robust against perturbations such as local measurements.
\end{abstract}

\date{\today}
\maketitle


\section{Introduction}
Quantum matter possesses a rich entanglement structure that awaits to be discovered. Significant effort has been devoted to studying bipartite entanglement~\cite{wangEntanglement2024,wangEntanglement2023,mao_sampling_2023, Braiorr-Orrs_Weyrauch_Rakov_2019,songExtracting2023,osterlohScaling2002,parez2023fermionic}, as quantified through measures such as entanglement entropy (EE) of a subregion and its complement. The EE has yielded numerous highly non-trivial insights about topological phases, and quantum critical theories both in and out of equilibrium~\cite{PasqualeCalabrese_2004,Amico2008,Hastings2010measure,humeniukQuantum2012,groverEntanglement2013,albaEntanglement2013,laflorencieQuantum2016,albaOut2017,Skinner2019,demidioEntanglement2020,zhaoScaling2022,songDeconfined2023,dengDiagnosing2024,zhangIntegral2023,zhouIncrmental2024}. 
However, entanglement can have more collective forms which cannot be detected via bipartite measures. A general question one can ask about a quantum many-body system is the following: are $m$ subgroups of spins (called parties) entangled with each other? A further interesting question follows: are the $m>2$ parties all involved in the entanglement?
If so, we say that the parties possess genuine multiparty entanglement (GME), the most powerful form of entanglement as it involves all the parties.
A major hurdle in answering these questions in realistic states is the computational difficulty in quantifying multipartite entanglement, whether genuine or not, as the number of spins grows. A promising avenue where tangible progress can be made is the {\it entanglement microscopy} approach, which aims to determine the entanglement between spins in small subregions. This has the added advantage that getting reliable data about the reduced density matrix of a microscopic subregion is less prone to finite-size effects since the correlation functions needed are more local. 
In pioneering work, the bipartite entanglement between two spins in the 1d transverse field Ising model (TFIM) has been determined by exploiting the exact solution~\cite{Giampaolo2013,Hofmann2014}. Later, this was extended to GME among 3 or 4 spins both in the groundstate and at finite temperature~\cite{Wen2024}. 
In 2d, where such exact solutions are not available, a powerful tool to study entanglement is quantum Monte Carlo (QMC). For instance, QMC has been used to compute Rényi negativity in 2d TFIM, revealing the short-range nature of bipartite entanglement at the QCP~\cite{Entanglement_Wu_2020}.
Building on these advancements, entanglement microscopy was recently extended to a broader class of models by employing QMC to directly sample the reduced density matrix (RDM) of subregions~\cite{wangEntanglement2024, mao_sampling_2023}.
For example, the full RDM of 2 and 3 spins in the 2d TFIM was obtained for large lattices. This provided new insights in this non-solvable model, which is a cornerstone in the study of magnetism and quantum criticality. First, the bipartite negativity between 2 spins was shown to vanish for non-adjacent sites, which offers a striking illustration of the short-ranged nature of entanglement in quantum matter~\cite{foe}. Second, GME for 3 adjacent spins in 2d was studied using a biseparability criterion, but was only detected at large values of $h$ far beyond the quantum critical point, and with a small magnitude. 

In this work, we further push the limits of entanglement microscopy in both the 1d and 2d TFIMs and extend it to 3d. This enables a comprehensive investigation of bipartite and multiparty microscopic entanglement in the TFIM as a function of the tuning parameter (transverse field $h$) and dimensionality. Here are the main conclusions:
\begin{itemize}
    \item Spins in a microscopic subregion share stronger GME in the paramagnetic phase compared to the ferromagnet. The GME is maximal near the QCP.
    \item The microscopic GME decreases rapidly with dimension, with the 1d TFIM having the strongest GME.
    \item When spins become disconnected, they rapidly lose GME. In 2d and 3d, we only find GME for adjacent sites, illustrating the short-ranged nature of multipartite entanglement in equilibrium quantum matter~\cite{foe}. For many disconnected subregions, we are able to prove biseparability, hence the vanishing of \emph{any} GME measure.
    \item The most collective GME, shared among all $N$ sites in the lattice, possesses a very simple expression in terms of an experimentally accessible observable: the magnetization parallel to the applied transverse field. This $N$-party GME decreases rapidly with dimension but is less robust (in relative terms) to local measurements in 1d compared to higher dimensions.
\end{itemize}

Many of the above results can be understood by invoking the {\it monogamy of entanglement}: as the number of neighbors increases in higher dimensions, the available entanglement must be shared more collectively among a larger number of parties, weakening the amount of entanglement that can be concentrated between any specific subset of spins.
Consequently, entanglement becomes more diluted across the system, leading to the observed decrease in both bipartite and multipartite entanglement.

In addition to the findings specific to Ising models, we have obtained general results regarding GME:
\begin{itemize}
    \item A completion of the previously incomplete proof that the biseparability criterion $I_2$ lower bounds a true measure of GME, the genuinely multipartite concurrence (GMC). 
    \item Symmetries of the RDM can significantly simplify the computation of $I_2$, though distinct structures in the optimal parameters persist.
    \item The explicit numerical evaluation of the GMC for mixed states via the convex roof extension. 
    \item The global GMC among all qubits in uniform (permutation invariant) pure states reduces to single-site properties.
\end{itemize}

Thus, by providing a comprehensive set of entanglement
measures across different dimensions, our study contributes the first
extensive dataset capturing GME for archetypal QCPs in 1d, 2d, and
3d systems.

\section{Quantum Ising Models and Entanglement Measures}

\begin{figure}[htp!]
\centering
\includegraphics[width=\columnwidth]{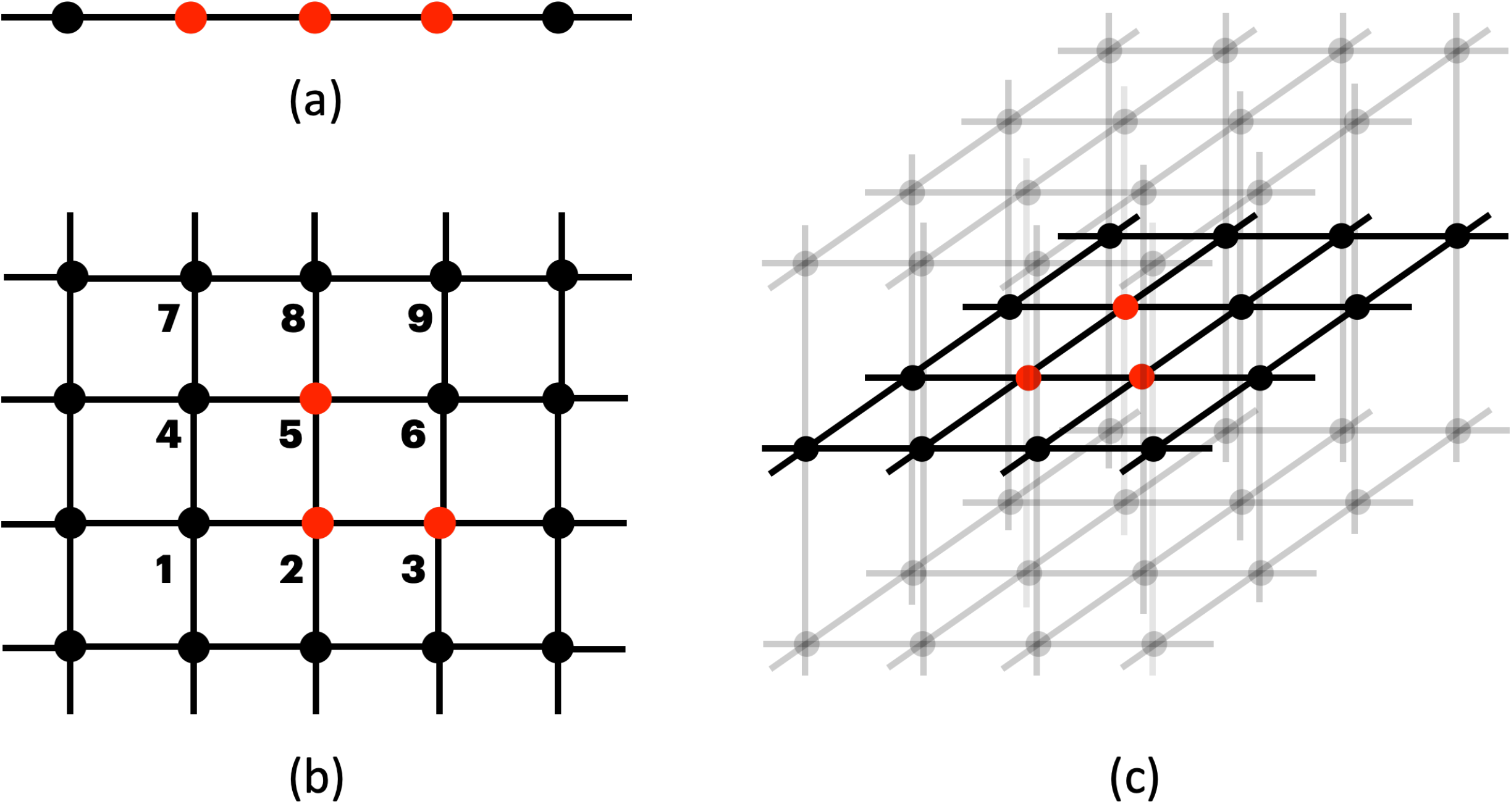}
\caption{\textbf{Subregions in 1, 2, and 3 dimensions.} Microscopic subregion $A$ (red dots, using 3 sites as an example) for the (a) $d=1$ (b) $d=2$ (c) $d=3$ systems. We investigate the bipartite and multipartite entanglement properties of such subregions near the QCPs of the quantum Ising models.}
\label{fig:fig1}
\end{figure}

We investigate the entanglement properties of TFIM in 1d, 2d, and 3d via the multipartite entanglement microscopy protocol~\cite{wangEntanglement2024}. We begin with the Hamiltonian of the ferromagnetic TFIM: 
\begin{equation}
    H=-J \sum_{\langle i, j\rangle} \sigma_i^z \sigma_j^z-h \sum_i \sigma_i^x,
\end{equation}
where $J = 1$ is the nearest-neighbor ferromagnetic exchange coupling and $h$ is the transverse field.
The QCPs correspond to $h_c = 1$~\cite{Pfeuty1970}, $h_c \approx 3.044$~\cite{hesselmannThermal2016}, and $h_c \approx 5.2$~\cite{Blote2002} for the 1d chain, 2d square lattice, and 3d cubic lattice, respectively.




\subsection{Bipartite entanglement}
\noindent{\textbf{Entanglement Entropy--}}The entanglement entropy reads
\begin{equation}
    \SvN=-\Tr\rho_A\log_2\rho_A
\end{equation}
where $\rho_A$ is the RDM of the subregion $A$. Note that we shall use the logarithm base 2 in this paper. 
In a pure state $|\Psi_{AB}\rangle$, $S^{\rm vN}(A)$ is a valid measure of entanglement between $A$ and its complement, $B$; it fails to be an entanglement measure for mixed states. In finite temperature QMC simulations, one usually scales the inverse temperature $\beta=1/T$ with the linear system size $L$ to reduce thermal effects and obtain the groundstate information. $\SvN$ of three adjacent sites are shown in Fig.~\ref{fig:fig2}.\\

\noindent{\textbf{Negativity---}}The logarithmic negativity is calculated with the RDM $\rho_A$ where region $A$ is subsequently divided into two parts, $A_1$ and $A_2$, and one takes the partial transpose with respect to $A_1$~\cite{zyczkowski98,Eisert98,VW02}, 
\begin{equation}
    \mathcal{E}=\log_2 \left\|\rho_A^{T_1}\right\|_1
\end{equation}
where the $|| \cdot ||_1$ is the matrix 1-norm, which equals the sum of the absolute eigenvalues of the partially transposed matrix $\rho_A^{T_1}$. The partial transpose of the RDM $\rho_A$ operate as  $\bra{\alpha_{1}\alpha_{2}}\rho_A^{T_1}\ket{\bar{\alpha}_{1}\bar{\alpha}_{2}}=\bra{\bar{\alpha}_{1}\alpha_{2}}\rho_A\ket{\alpha_{1}\bar{\alpha}_{2}}$;
the subscript denotes the subregion, $A_1$ or $A_2$.
As shown in Figs.~\ref{fig:fig3} and \ref{fig:fig4} below, we computed $\mathcal{E}$ for $A_1$ and $A_2$, where each subregion $A_i$ consists of either 1 site or 2 sites, across the QCPs in 1d, 2d, and 3d, respectively. \\



\subsection{Multipartite entanglement}\label{sec:GME}
Consider a $n$-partite quantum state $\rho$ on $\mathcal H_1\otimes\cdots\otimes\mathcal H_n$. There are $(2^{n-1}-1)$ number of ways to partition it into two subsystems. We denote the set of bipartitions by $\gamma= \{ \gamma_i \!=\! (A_i \vert B_i)\vert i=1, \ldots, 2^{n-1}-1\}$.
A biseparable pure state with respect to bipartition $A_i \vert B_i$ takes the form $\vert \phi^{\mathrm{bs}}_{A_i \vert B_i}\rangle=|\alpha\rangle_{A_i} \otimes|\beta\rangle_{B_i}$.
More generally, a biseparable mixed state is defined as a convex combination of biseparable pure states:
$\rho^{\mathrm{bs}}=\sum_k p_k\left|\phi_k^{\mathrm{bs}}\right\rangle\left\langle\phi_k^{\mathrm{bs}}\right|$
where $\left|\phi_k^{\mathrm{bs}}\right\rangle$ may be biseparable with respect to different partitions of the $n$ parties~\cite{Guhne2009}.
A state $\rho$ that is not biseparable is called \textit{genuinely multipartite entangled}~(GME). 

One measure of genuine multipartite entanglement that is zero for all biseparable states and positive for all GME states is the GME-concurrence (GMC)~\cite{Ma2011}. 
For a pure state $|\Psi\rangle$, the GME-concurrence is the minimum bipartite concurrence for all bipartitions of the system:  
\begin{align} \label{eq:gmc}
C_{\rm GME}(|\Psi\rangle)=\min _{\gamma_i \in \gamma} \sqrt{2[1-\operatorname{Tr}(\rho_{A_i}^2)]}
\end{align}
where $\rho_{A_i} = \Tr_{B_i}(\vert\Psi\rangle \langle \Psi \vert )$ for bipartition $\gamma_i = (A_i \vert B_i)$.
For a mixed state $\rho$,  $C_{\rm GME}(\rho)$ is defined through a standard convex roof extension by minimizing over all mixtures of pure states, $\rho=\sum_k p_k |\Psi_k\rangle\langle\Psi_k|$:
\begin{align}
    C_{\rm GME}(\rho) = \min_{\{p,\Psi\} } \sum_k p_k C_{\rm GME}(|\Psi_k\rangle)
\end{align} 
Convex roof extensions are in general challenging to optimize; however, Ref.~\cite{Ma2011} proves that there exists a more accessible lower bound for $C_{\rm GME}$, the $I_2$ criterion. $I_2$ is defined through the following maximization~\cite{Huber2010}
\begin{equation}
    I_2(\rho)=\max _{| \Phi \rangle}\left\{\sqrt{\langle\Phi| \rho^{\otimes 2} \Pi|\Phi\rangle}-\sum_i \sqrt{\langle\Phi| \mathcal{P}_i^\dagger \rho^{\otimes 2} \mathcal{P}_i|\Phi\rangle}\right\}
    \label{eq:eq4}
\end{equation}
where the objective function to be maximized is evaluated by considering two copies of the state $\rho^{\otimes 2}$ on the doubled Hilbert space $(\mathcal H_1\otimes\cdots\otimes\mathcal H_n)^{\otimes 2}$, on which we maximize over all product states 
$\vert \Phi \rangle = |\varphi_1\rangle \otimes|\varphi_2\rangle \otimes \ldots \otimes|\varphi_{2n}\rangle$.
$\Pi$ is the global swap operator that exchanges the two copies, while $\mathcal{P}_i=\Pi_{A_i} \otimes \mathbbm{1}_{B_i}$ is the swap operator for bipartition $\gamma_i=(A_i\vert B_i)$ which exchanges part $A_i$ of the two copies. 
If $I_2 > 0$, there is GME, while no conclusion can be made if $I_2 \leq 0$. 
Although $I_2$ is weaker than the GMC, it has one important advantage (in addition to lower computational cost): it is obtained by maximization. So even if one finds a non-optimal value that is strictly positive, the state is guaranteed to be GME. In contrast, one cannot rigorously establish that a finite GMC obtained through a numerical convex roof optimization will not be supplanted by a global minimum with $C_{\rm GME}=0$. In practice, we can perform checks on the GMC to gain confidence about the values obtained through the minimization.

It is important to clarify, however, that the lower bound in Ref.~\cite{Ma2011} was only proved for $I_2$ optimized by a restrictive set of $\ket{\Phi}$.
If we define the objective function to be $I_2[\rho,\ket{\Phi}]$, then the lower bound $2 I_2[\rho,\ket{\Phi}] \leq C_{\text{GME}}(\rho)$ was proved for $\ket{\Phi} = U \otimes U \ket{00\ldots01\ldots11}$ where $U$ is a local unitary matrix. 
In Appendix Sec.~\ref{Appendix:lower_bound_proof}, we complete this proof by showing that the lower bound holds for all product states $\ket{\Phi}$.
Under the restricted parametrization, \( I_2 \) reduces to another GME criterion \( W_1 \)~\cite{Guhne_2010}, see Appendix Sec.~\ref{Appendix:I2 stronger than W1} for a derivation. For three spins, $W_1$ is defined as
\begin{equation}
W_1 = \max_{\mathrm{LU}} \left( |\rho_{1,8}| - \sqrt{\rho_{2,2} \rho_{7,7}} - \sqrt{\rho_{3,3} \rho_{6,6}} - \sqrt{\rho_{4,4} \rho_{5,5}} \right),    
\end{equation}
where \( \rho_{i,j} \) denotes matrix elements of the three-site RDM, and the maximization is performed over local unitary (LU) transformations \( \rho \rightarrow U \rho U^{\dagger} \) with \( U = U_1 \otimes U_2 \otimes U_3 \). The criterion for multipartite entanglement is $W_1 > 0$.
Empirically, we previously employed both \( W_1 \) and a related criterion \( W_2 \)~\cite{Guhne_2010} to detect GME in 1d and 2d systems~\cite{wangEntanglement2024}. While both \( W_1 \) and \( W_2 \) successfully detected GME in 1d, they failed to do so in 2d. In contrast, the \( I_2 \) criterion successfully detects GME across 1d, 2d, and 3d, covering both three- and four-spin configurations, see Table~\ref{table:GME_concurrence} for a comparison of $C_{\rm GME}$, \( I_2 \) and $W_1$ for three adjacent spins in 1d, 2d and 3d. 
$I_2$ has also been successfully applied to detect GME in other 1d models, such as the XY Ising chain~\cite{Giampaolo2013} and the cluster-Ising model~\cite{giampaoloGenuine2014}. These results, combined with its performance across 1d, 2d, and 3d for both exact diagonalization (ED) and QMC states, highlight the robustness of  $I_2$  as a versatile tool for quantifying GME in complex quantum systems.

\subsection{Numerical tomography}
To simulate the TFIM, we adopt the stochastic series expansion (SSE) QMC technique~\cite{wangEntanglement2024,sandvik2019stochastic,zhaoHigher2021} and keep the inverse temperature $\beta=1/T=L$ for 1d and 2d while $\beta=1/T=2L$ for 3d to approach the groundstate RDM. The simulated linear system sizes are up to $L=48$ for 1d, $L=24$ for 2d, and $L=8$ for 3d lattices. The entanglement regions are made of 2, 3 (as shown in Fig.~\ref{fig:fig1}) and 4 nearby sites. We complement the QMC results with analytic solutions in 1d via Jordan-Wigner transformation, exact diagonalization (ED) in 2d with $5\times 5$ and $6\times 5$, and 3d with $L=3$. 

Based on the above definitions, we now present the ED and QMC results accordingly. The obtained RDMs for the 1d, 2d, and 3d TFIMs are also given in Appendix Sec.~\ref{Appendix:RDM_matrox} and on the \href{https://github.com/songmengh/Ising_RDMs/tree/main}{GitHub repository}~\cite{RDM_git}, such that interested readers can carry out the independent entanglement measurements and make comparisons with those presented here.

\section{Bipartite Entanglement}

\subsection{Entanglement entropy}

\begin{figure}[htp!]
\centering
\includegraphics[width=\columnwidth]{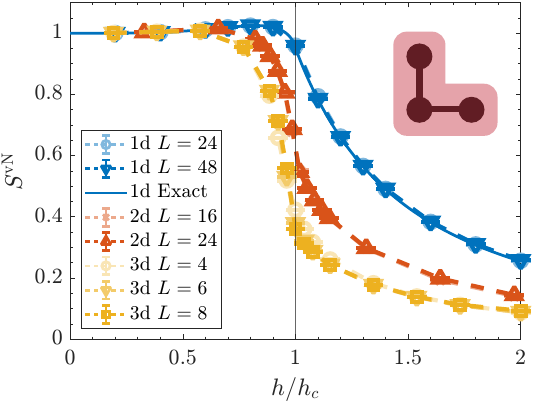}
\caption{\textbf{Entanglement entropy $\SvN$ of 3 adjacent sites in the TFIM.} The inset shows the entanglement region $A$ for 2d and 3d cases. In the 1d case, the $\SvN$ are computed with 3 adjacent sites as in Fig.~\ref{fig:fig1} (a).}
\label{fig:fig2}
\end{figure}

As a warm-up, we first compute the EE $\SvN$ across the QCPs of 1d, 2d and 3d TFIM, as shown in Fig.~\ref{fig:fig2}. In the 1d case, we consider the entanglement subregion with 3 adjacent sites and find the $L=24$ and $48$ sites chain give converged results with the exact solution at the thermodynamic limit~\cite{osterlohScaling2002}. For the 2d and 3d cases, the 3-site subregion is shown in the inset of Fig.~\ref{fig:fig2}, and similar convergence of \( \SvN \) is observed with increasing $L$.

To gain further insights into the behavior of \( \SvN \) near the QCP, we examine the simpler case of a single-site subregion.
For subregion $A$ containing only 1 site, due to symmetry, the $2\times 2$ RDM is parametrized by the transverse field magnetization $\left\langle\sigma^x\right\rangle$ only: $\rho_1=\frac{1}{2}(I+\left\langle\sigma^x\right\rangle \sigma^x)$. 
As explained in our previous work~\cite{wangEntanglement2024}, near the QCP, the scaling behavior of $d \SvN/dh$ can be derived from the critical exponents of $\left\langle\sigma^x\right\rangle$ obtained from field theory. That is,
\begin{align}
    \frac{d\SvN}{dh} &\sim \log_2|h-h_c| \  (\text{1d, 3d}) \label{eq:entEntropyScaling_1} \\    
    \frac{d\SvN}{dh} &\sim |h-h_c|^{-0.11}, \  (\text{2d}) 
    \label{eq:entEntropyScaling_2}
\end{align}
with powers determined by the scaling dimension at the corresponding QCP, i.e., $\Delta_\varepsilon \nu$. 
In 1d, the logarithmic divergence of $dS^{\rm vN}/dh$ can be derived explicitly from the exact solution. In 2d, where an exact solution is unavailable, we analyze the scaling behavior of $\av{\sigma_x}$, computed as the off-diagonal elements of our sampled RDM. The results, shown in Fig.~\ref{fig:2d_sigma_x} in Appendix Sec.~\ref{Appendix:Singular Scaling}, agree with prior work~\cite{xie2012coarse}, yielding consistent critical exponents with $\Delta_\epsilon\nu\approx 0.89$, thereby supporting the expected scaling $dS^{\rm vN} / d h \sim \left|h-h_c\right|^{\Delta_{\varepsilon} \nu-1}$.

\subsection{Logarithmic negativity}

\begin{figure}[htp!]
\centering
\includegraphics[width=\columnwidth]{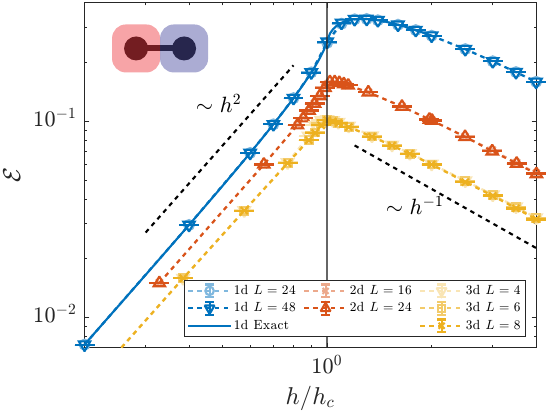}
\caption{\textbf{Log negativity of two adjacent sites.} 
The logarithmic negativity scales as  $h^2$  at small fields and  $h^{-1}$  at large fields for  $d=1, 2,$  and  $3$ . Singularities are observed at the QCP in all dimensions. The inset illustrates the geometry of the 2-site entanglement region  $A$, with shaded areas indicating the 1-site partitions  $A_1$  and  $A_2$. The plot also demonstrates convergence of the results with increasing system size  $L$.
}
\label{fig:fig3}
\end{figure}

\begin{figure}[htp!]
\centering
\includegraphics[width=1.02\columnwidth]{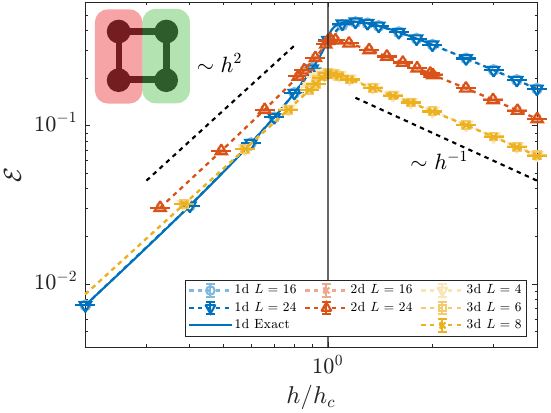}
\caption{\textbf{Log negativity of 4 adjacent sites.} 
The logarithmic negativity scales as  $h^2$  at small fields and  $h^{-1}$  at large fields for  $d=1, 2,$  and  $3$ . Singularities are observed at the QCP in all dimensions. The inset illustrates the geometry of the 4-site entanglement region  $A$, with shaded areas indicating the 2-site partitions  $A_1$  and  $A_2$. The plot also demonstrates convergence of the results with increasing system size $L$.
}
\label{fig:fig4}
\end{figure}

The logarithmic negativity for adjacent subregions  $A_1$  and  $A_2$  is shown for 1-site subregions in Fig.~\ref{fig:fig3} and 2-site subregions in Fig.~\ref{fig:fig4}, based on exact solutions for 1d and QMC simulations for 1d, 2d, and 3d systems.

In Figs.\ref{fig:fig3} and \ref{fig:fig4}, the logarithmic negativity  $\mathcal{E}$  is shown for adjacent subregions with  $A_1 = A_2 =$ 1  site and 2 sites, respectively. Across 1d, 2d, and 3d QCP, $\mathcal{E}$ exhibits pronounced peaks near the QCP, accompanied by clear singularities. 
However, the magnitude of these peaks diminishes as the spatial dimension increases. This trend reflects the monogamous nature of entanglement, where in higher dimensions, a subregion must share its entanglement with a larger number of neighboring subregions, thereby reducing the amount distributed to each. 
This behavior is consistent with previous analytical results~\cite{Ou2007, Coffman2000}, and, to the best of our knowledge, provides the first unbiased demonstration of this property in the QCPs of a quantum many-body system.


For all cases, the groundstate $\eps$ vanishes in the two limits $h\rightarrow 0$ $(\infty)$, and is finite otherwise. This is in contrast to what happens in the presence of a finite temperature, where there is a ``sudden-death'' phenomenon: $\eps$ drops to zero at a finite $T$~\cite{wangEntanglement2024,javanmardSharp2018}. A sudden death also occurs as a function of the separation between the spins, as we discuss below.
At both $h=0$ and $h=\infty$, for both two spins (Fig.~\ref{fig:fig3}) and two pairs of spins (Fig.~\ref{fig:fig4}), the groundstates are separable and non-full rank. This means that they lie on the boundary of the separable continent~\cite{foe, Wen2024, Pittenger2002} and thus no sudden death of negativity occurs as a function of the field $h$.

In 1d, using exact solutions for the two-spin correlation functions~\cite{Pfeuty1970}, we can extract the asymptotic behavior of $\eps$ for two nearest neighboring spins at small field $\eps \sim \frac{1}{8} h^2$ and at large field $\eps \sim \frac{1}{2} h^{-1}$, which agrees with Fig.~\ref{fig:fig3}. 

Similar scaling behavior is observed in 2d and 3d, as shown in Fig.~\ref{fig:fig3} and \ref{fig:fig4}, where  $\mathcal{E} \sim h^2$  at small fields and  $\mathcal{E} \sim h^{-1}$  at large fields for both two-spin and two-pair configurations, 
which can be heuristically understood as follows: the negativity depends on off-diagonal matrix elements of the reduced density matrix $\rho_A$, which arise from processes involving simultaneous spin flips in the two subsystems. 
At small $h$ the perturbation is $\delta H=-h\sum X_i$, so such processes occur at second order in perturbation theory, leading to a quadratic scaling with $h$. 
Conversely, at large $h$ the perturbation is $\delta H=-h^{-1} \sum Z_iZ_j$, so the correction occurs at first order leading to $1 / h$ scaling. The consistency across dimensions highlights the robustness of these asymptotic behaviors.


\section{Microscopic multiparty entanglement}
We now shift our focus to multipartite entanglement.
To quantify GME, we evaluate three measures \( C_{\text{GME}} \), \( I_2 \), and \( W_1 \) for three adjacent spins at the QCP are summarized in Table~\ref{table:GME_concurrence}. 
The table provides a comparative perspective across 1d, 2d, and 3d using both ED and QMC. While \( C_{\text{GME}} \) represents the most rigorous measure of GME, the results confirm that \( I_2 \) reliably detects GME across all dimensions and computational methods. Notably, \( W_1 \), being less sensitive, fails to detect GME in 2d and 3d, while \( I_2 \) remains positive even as the values decrease significantly with increasing dimensionality. Additionally, we observe that \( 2I_2 \), which provides a lower bound for \( C_{\text{GME}} \) as discussed earlier, is generally loose, especially in higher dimensions, where the difference between \( 2I_2 \) and \( C_{\text{GME}} \) becomes more pronounced. 
For the estimate of \( C_{\text{GME}} \), we perform the optimization over convex roof extensions using the Matlab library LibCreme~\cite{ROTHLISBERGER2012155}. 
Practically, the optimization of $C_{\mathrm{GME}}$ yields only an upper bound that cannot certify GME when the value approaches zero. In contrast, $I_2$, as a lower bound, certifies GME whenever positive. Therefore, we will now focus on $I_2$ as a measure of tripartite and four-partite GME.

\begin{table}[h!]
\centering
\renewcommand{\arraystretch}{1.5} 
\begin{tabular}{lccccc}
\toprule
\textbf{\quad} & Lattice & \( h \) & \( C_{\text{GME}} \) & \( I_2 \) & \( W_1 \) \\
\midrule
\multirow{1}{*}{1d} & \( L = \infty \) & \( 1 \) & \( 0.186 \) & \( 0.04248 \) & \( +0.035 \) \\
\midrule
\multirow{2}{*}{2d} & \( L = 5 \) (ED) & \( 3 \) & \( 0.081 \) & \( 0.00551 \) & \( -0.012 \) \\ 
                    & \( L = 24 \) (QMC) & \( 3.044 \) & \( 0.076 \) & \( 0.00423 \) & \( -0.019 \) \\
\midrule
\multirow{2}{*}{3d} & \( L = 3 \) (ED) & \( 5.2 \) & \( 0.054 \) & \( 0.00164 \) & \( -0.010 \) \\ 
                    & \( L = 8 \) (QMC) & \( 5.2 \) & \( 0.048 \) & \( 0.00092 \) & \( -0.011 \) \\
\bottomrule
\end{tabular}
\caption{The biseparability criterion \( I_2 \), our estimate for the true measure of GME \( C_{\text{GME}} \), and \( W_1 \) for the three most adjacent spins at the QCP from both ED and QMC. \( C_{\text{GME}} \) is obtained by minimizing over all \( K \)-component pure-state decompositions \(\{ p_i, |\phi_i\rangle \, | \, i = 1, \dots, K \}\) of the RDM; here, we use \( K \leq 18 \).}
\label{table:GME_concurrence}
\end{table}

\begin{figure}[htp!]
\centering
\includegraphics[width=\columnwidth]{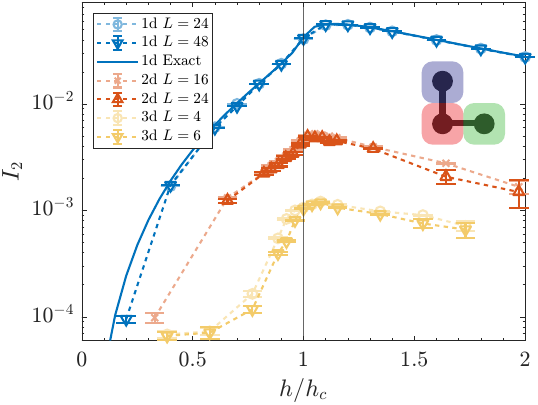}
\caption{\textbf{Entanglement criterion $I_2$ for 3 spins.} The 3-party $I_2$ criterion detects genuine entanglement between 3 neighboring spins in 1d, 2d, 3d. Different linear sizes $L$ are shown to illustrate convergence. We note that 1) the peak occurs in the paramagnetic phase near the QCP, and 2) $I_2$ decreases rapidly with the dimension $d$.  
}
\label{fig:fig5}
\end{figure}

Using $I_2$, we now analyze the tripartite entanglement in subregion $A$ composed of 3 adjacent sites across different spatial dimensions. The results are shown in Fig.~\ref{fig:fig5}.
For each spatial dimension, our results demonstrate clear convergence with increasing linear system size $L$.
In the 1d case, the converged data also agrees with the exact solution in the thermodynamic limit, shown as the solid blue line.
Across all dimensions, $I_2$ detects GME in a wide range of $h$ values and exhibits a peak in the paramagnetic phase near the QCP. The magnitude of $I_2$, however, decreases significantly with increasing $d$, consistent with the monogamous nature of entanglement.
This contrasts with $W_1$ and $W_2$, which fail to detect GME near the QCP in 2d and 3d~\cite{wangEntanglement2024}, as discussed earlier. To validate our QMC results, we complement them with exact diagonalization for $5 \times 5$, $6\times 5$ and $3 \times 3 \times 3$ lattices, showing agreement with the QMC data (Appendix Sec.~\ref{Appendix:Singular Scaling}).

\begin{figure}[htp!]
\centering
\includegraphics[width=\columnwidth]{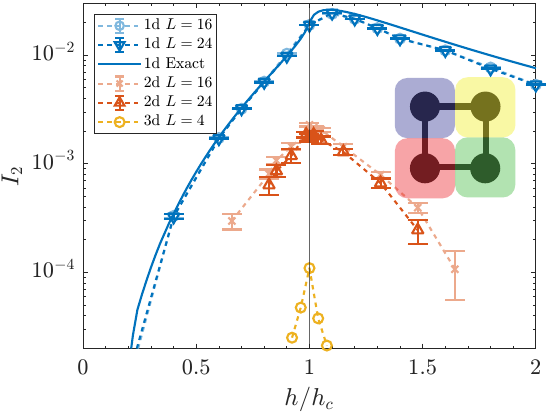}
\caption{\textbf{Entanglement criterion $I_2$ for 4 spins.} Four-party $I_2$ for 4 nearest sites as a function of field $h$ in 1d, 2d, and 3d. } 
\label{fig:fig6}
\end{figure}

Finally, we turn to a subregion $A$ composed of the 4 closest sites, a line in 1d and a square plaquette in 2d / 3d. We investigate 4-party GME via $I_2$ as a function of $h$, as shown in Fig.~\ref{fig:fig6}. We observe the same trends as for the 3-site case. Overall the 4-site $I_2$ values are lesser than the 3-site results in Fig.~\ref{fig:fig5}, which is expected since the number of nearest neighbors of subregion $A$ is larger for the plaquette compared to the 3-site geometry. In 3d, the value is very small, $O(10^{-4})$, which makes it difficult to resolve 
since the error bars of the RDM matrix elements are of the same order. Our exact diagonalization results on $3\times3\times3$ corroborate the finite but small values of $I_2$ near the QCP. We note that our ED results are strictly greater than the QMC ones for larger $L$ both in 2d and 3d, consistent with the observation that entanglement is stronger when the environment, $B$, is smaller. 



We now analyze the asymptotic scaling of $I_2$ at small and large fields. 
Unlike negativity,  $I_2$  is non-linear in the RDM and involves an optimization, making its field dependence more intricate.
Nevertheless, we can perform a similar scaling analysis by noting that GME measures depend on off-diagonal elements of the reduced density matrix that involve simultaneous spin flips in all subsystems.
At small field, the perturbation analysis predicts  $h^3$  and  $h^4$  scaling for tripartite and four-partite entanglement, respectively, while at large field,  $h^{-2}$  scaling is expected for both.
Empirically, for the 1d case, as shown in Fig.~\ref{fig:I2_vs_h_power} in the Appendix, all these exponents agree with the data except for the large-field four-partite case, which exhibits  $h^{-3}$. 
This deviation highlights that the scaling analysis provides a lower bound on the decay rates, as cancellations in the perturbative corrections can lead to faster decay. 
Similar behavior is expected in higher dimensions, though current QMC data do not resolve these regimes.


\subsection{Singular scaling near the QCP}\label{section:singular_scaling}

As previously demonstrated~\cite{wangEntanglement2024}, the derivative of a generic entanglement measure exhibits a divergent behavior near the QCP. This applies even to entanglement measures that rely on an optimization procedure, such as $I_2$. Specifically, the derivative of any entanglement measure $\mathcal{M}$ scales as $d \mathcal{M}/d h =\alpha| h-\left.h_c\right|^{\Delta_{\varepsilon} \nu-1}$ when $h \rightarrow h_c$. One application of this general scaling in bipartite entanglement is already seen in Eq.~\ref{eq:entEntropyScaling_1} and \ref{eq:entEntropyScaling_2} for entanglement entropy. The same scaling holds for the Log Negativity, as shown in Fig.~\ref{fig:LN_log_scale} in the Appendix Sec.~\ref{Appendix:Singular Scaling}.

Applying this scaling for GME, in 1d, $I_2$ shows a logarithmic divergence
\begin{equation}
    \left. \frac{d I_2}{d h} \right|_{h \rightarrow h_c} = -\alpha \log |h - h_c| + \beta
\end{equation}
where $\alpha = 0.11,0.07$ for three and four spins. The plot of $d I_2 / d h$ versus $h/hc$ and $\log |h - h_c|$ is shown in  Fig.~\ref{fig:I2_1d_logscale} in the Appendix Sec.~\ref{Appendix:Singular Scaling}.
The 1d case was previously analyzed in Ref.~\cite{Giampaolo2013}, but they find different values of $I_2$ compared to us. For example, at the QCP, they find $I_2 = 0.07$, while we find $0.043$. 
The discrepancy in \( I_2 \) values arises from two key differences. 
First, their $I_2$ was multiplied by a factor of 2, and they optimized only over local unitary transformations on the reduced density matrix $\rho \rightarrow U^\dagger \rho U$, reducing the criterion to $W_1$, as explained in Sec.~\ref{sec:GME}.
In contrast, our $I_2$ is optimized over all product states $\vert \Phi\rangle$, making it a stronger criterion for detecting GME, see Appendix Section \ref{Appendix:I2 stronger than W1} for proof. 
As shown in Appendix Fig.~\ref{fig:I2_W1_comparison} , their $I_2/2$ aligns exactly with our $W_1$, indicating that their optimization effectively implements the $W_1$ criterion.
They also find a different coefficient, $\alpha = 0.78$, which is approximately four times the value we find for $2W_1$ in our previous work~\cite{wangEntanglement2024}.
Even after accounting for the aforementioned differences, some discrepancies remain unexplained.
The logarithmic scaling is also seen in another GME measure~\cite{Hofmann2014}.
Using the scaling dimensions, we expect the same logarithmic scaling in 3d, and a power-law scaling $d I_2/d h \sim \left| h - h_c \right|^{-0.11}$ in 2d. Due to insufficient data resolution, we cannot extract this scaling dimension from our QMC results, but the derivative becomes maximal near the QCP, see Fig.~\ref{fig:I2_vs_h_2d_TFIM},~\ref{fig:I2_vs_h_3d_TFIM} in the Appendix Sec.~\ref{Appendix:Singular Scaling}.


\subsection{Decrease of entanglement with separation}

\begin{table}[t]
    \centering
    \renewcommand{\arraystretch}{1.5} 
    \begin{tabular}{lccc}
        \toprule
        & \multicolumn{3}{c}{\( I_2 \times 10^3 \)} \\
        Subregion & $5\times5$ (ED) & $6\times5$ (ED) & $24\times24$ (QMC) \\
        \midrule
        \textbf{124}  & \( +5.51 \) & \( +5.30 \) & \( +4.9 \) \\
        \textbf{123}  & \( +4.60 \) & \( +4.00 \) &  \\
        126           & \( -0.27 \) & \( -0.28 \) &  \\
        146$^{\rm bs}$          & \( -0.31 \) & \( -0.32 \) & \( -0.3 \) \\
        129$^{\rm bs}$         & \( -0.33 \) & \( -0.33 \) &  \\
        153$^{\rm bs}$           & \( -0.39 \) & \( -0.40 \) &  \\
        137$^{\rm bs}$           & \( -0.43 \) & \( -0.44 \) &  \\
        \bottomrule
    \end{tabular}

    \vspace{0.3cm} 

    \begin{tabular}{lccc}
        \toprule
        & \multicolumn{3}{c}{\( I_2 \times 10^3 \)} \\
        Subregion & $5\times5$ (ED) & $6\times5$ (ED) & $24\times24$ (QMC) \\
        \midrule
        \textbf{1245}  & \( +3.026 \) & \( +2.896 \) & \( +2.55 \) \\
        \textbf{1256}  & \( +0.227 \) & \( +0.089 \) &  \\
        \textbf{1236}  & \( +0.097 \) & \( -0.033 \) &  \\
        \textbf{1235}  & \( -0.010 \) & \( -0.050 \) &  \\
        1453      & \( -0.034 \) & \( -0.036 \) &  \\
        1436         & \( -0.034 \) & \( -0.035 \) & \( -0.07 \) \\
        1269           & \( -0.034 \) & \( -0.034 \) &  \\
        2567           & \( -0.040 \) & \( -0.041 \) &  \\
        2468$^{\rm bs}$           & \( -0.057 \) & \( -0.053 \) &  \\
        1379$^{\rm bs}$           & \( -0.053 \) &  \( -0.056 \)&  \\
        \bottomrule
    \end{tabular}
    \caption{\( I_2\) values for three- and four-spin subregions from exact diagonalization (ED) on \( 5 \times 5 \) and \( 6\times5 \) lattices at \( h = 3 \), and quantum Monte Carlo (QMC) on a \( 24 \times 24 \) lattice at \( h = 3.044 \). All values are multiplied by \( 10^3 \) for clarity. Connected subregions are indicated in bold. The superscript `bs' means the state is biseparable (hence without GME), as certified by an algorithm introduced in Ref.~\cite{Kampermann2012}. The site ordering on the $L_x\times L_y$ square lattice  reads (the $x$-direction is horizontal): 
    \(\begin{smallmatrix}
        7 & 8 & 9 \\
        4 & 5 & 6 \\
        1 & 2 & 3
    \end{smallmatrix}\).}
    \label{table:I2}
\end{table}

Finally, we analyze the decrease of both bipartite and genuine multipartite entanglement as the separation between the subregions $A_i$ increases at the QCP. Bipartite entanglement, as measured by negativity, exhibits a rapid decline with separation characterized by the sudden death distance $r_{\mathrm{sd}}$, marking the point where negativity drops to zero. 
Here the distance is defined as the minimum distance between a site in subregion $A_1$ and another in $A_2$.
Let us begin with the case where each party is made of a single site, which we call the $(1,1)$ configuration, as shown in the inset of Fig.~\ref{fig:fig3} for adjacent sites ($r=1$). Using the exact solution we find that $r^{\text{1d}}_{\text{sd}}=3$ in 1d, in agreement with Ref.~\cite{javanmardSharp2018}.
In 2d, the QMC analysis shows that this distance decreases to $r^{\text{2d}}_{\text{sd}}=\sqrt{2}$, while ED on the $5\times5$ and $6\times 5$ lattices have $r^{\text{2d}}_{\text{sd}}=2$. 
In 3d, our QMC results yield $r^{\text{3d}}_{\text{sd}}=\sqrt{2}$, consistent with the trend observed in lower dimensions. In the $(2,2)$ configuration, depicted in Fig.~\ref{fig:fig4}, the sudden death distance increases to $r^{\text{1d}}_{\text{sd}}=4$ in 1d~\cite{javanmardSharp2018}. 
In 2d, we consider various separations of two parallel pairs. Using $5\times5$, $6\times5$ ED at $h=3$ and $24\times24$ QMC at $h=3.044$, we get respectively $\eps(1)=0.341, 0.335, 0.344$, $\eps(2)=0.00886, 0, 0.0001$ and $\eps(3)=0$ ($6 \times 5$ ED), showing a rapid sudden death of the LN (the QMC error bars at $r=2$ include zero). We go further and ask whether the vanishing of the LN also coincides with the vanishing of bipartite entanglement between the two pairs. To do so we evaluate the geometric entanglement~\cite{vedral1997quantifying}
\begin{align}
    \mathcal D(\rho)=\mbox{min}_{\rho_{\rm sep}}\sqrt{\Tr(\rho-\rho_{\rm sep})^2} \label{eq:geometric_distance}
\end{align}
which is the smallest distance with respect to the Frobenius norm, i.e.\/ the Hilbert-Schmidt metric, between the separable set defined by mixtures $\rho_{\rm sep}=\sum_k p_k \rho_1^k\otimes\rho_2^k$, and the density matrix $\rho$; here $\rho^k_i$ is a valid density matrix for the $i$th party ($i=1$ or 2). 
By performing the optimization over the separable states using the Gilbert algorithm~\cite{Pandya2020, Gilbert1966}, we find the following upper bounds on $\mathcal{D}$
for the RDM from $6\times5$ ED: $\mathcal{D}(r\!=\!1) = 0.10$, $\mathcal{D}(r\!=\!2) = 2.3 \times 10^{-4}$, and $\mathcal{D}(r\!=\!3) = 2.1 \times 10^{-4}$\footnote{Ref.~\cite{Pandya2020} choses random product states at each iteration to update the separable state. In contrast, we follow the original deterministic protocol of Gilbert~\cite{Gilbert1966}, which performs an optimization at each iteration to find a product state, leading to faster convergence.}. 
We found that the resulting approximate closest separable state does not yield an entanglement witness, so these values are not fully converged, and can be further reduced with additional work.
Nevertheless, the small distances suggest that any bound entanglement for $r\geq 2$ would be very weak, if present.

For GME, Table.~\ref{table:I2} summarizes $I_2$ for various three- and four-spin subregions. 
Our findings indicate that $I_2$ becomes negative 
when the subregions are disjoint, signifying the absence of GME between disconnected parts. In 1d, GME is detected by $I_2$ only for adjacent subregions, and this pattern continues in 2d, where no GME is detected for any disjoint subregion. 
In 3d, using ED $3\times 3\times 3$, we found no GME in disjoint subregion 126 (see indices in Table~\ref{table:I2}) and other subregions which involve out-of-plane sites. 
For the less compact subregions 146, 129, 153, 137, 2468, and 1379, we certify their biseparability using the iterative algorithm introduced in Ref.~\cite{Kampermann2012}. A local filtering transformation is applied before the algorithm to maximize the minimum eigenvalue of the RDM.
For subregion 126, where \( I_2 \) does not detect GME, we estimated GMC $C_{\rm GME} = 1.7 \times 10^{-3}$ and $1.1\times 10^{-3}$ for $5 \times 5$ and $6 \times 5$ ED, respectively. 
Benchmarking our optimization for the GMC on various states suggests that these values are consistent with a biseparable state.
 
We performed an additional check by calculating an upper bound on the geometric distance to the biseparable set, \( \mathcal{D}_{\text{biSEP}}(\rho) \), defined analogously to Eq.~\eqref{eq:geometric_distance}, as a measure for GME. For the RDM from \( 5 \times 5 \) and \( 6 \times 5 \) ED, we obtained  the following upper bounds \( \mathcal{D}_{\text{biSEP}} = 2.20 \times 10^{-4} \) and \( 2.09 \times 10^{-4} \), respectively. These bounds suggest that any GME for disconnected subregions would be extremely weak, if present.
In all dimensions, we see a consistent trend of rapidly decreasing GME with separation.

We note that in Ref.~\cite{Hofmann2014}, GME was detected in the 1d Ising model for both connected subregions, such as 123 and 1234, and disjoint subregions, including 124, 1245, and 1235. This detection was based on an alternative GME measure involving the optimization of an entanglement witness using semidefinite programming (SDP)~\cite{Ghune2011}.
However, the detected GME for disjoint subregions was more than ten times smaller than that of the connected subregions, and no GME was detected for subregions 135 and 125, which are separated by two spins. 
Additionally, they used an iterative algorithm to show that subregions 135 and 125 are biseparable.  
This finding highlights weak GME in some disjoint subregions in 1d, although $I_2$ cannot detect it. To further shed light on the disjoint GME, we evaluated the GMC in the 1d critical Ising chain for subregions 123 and 124, 
$C_{\rm GME} = 0.1857$, $0.0122$. We see that separating the third site from the pair 12 leads to a decrease by more than an order of magnitude. For subregion 135 we confirmed its biseparability via the iterative algorithm preceded with a filtering procedure~\cite{Hofmann2014, Kampermann2012}.

\subsection{Symmetries of Optimal Product States}\label{Appendix:Symmetry_I2}

The symmetries of the RDM play a central role in shaping the landscape of the objective function for \(I_2\). Given that the RDM of the TFIM is a real matrix, \(\ket{\Phi}\) and its complex conjugate should give the same value for $I_2$. 
Empirically, in all dimensions, we observe that results obtained using the full complex parametrization are identical to those obtained by restricting \(\ket{\Phi}\) to a real product state, expressed as:
\[
\ket{\Phi} =  \left( \begin{array}{c} \cos(\theta_1) \\ \sin(\theta_1) \end{array} \right)
\otimes
\left( \begin{array}{c} \cos(\theta_2) \\ \sin(\theta_2) \end{array} \right)
\otimes
\cdots
\otimes
\left( \begin{array}{c} \cos(\theta_n) \\ \sin(\theta_n) \end{array} \right)
\]
We note that not all real RDMs have a real optimal product state, so the above observation, although reasonable, is non-trivial.

The invariance of the RDM under unitary transformations further shapes the optimization landscape. If the RDM is invariant under \(\rho \to U^\dagger \rho U\), the objective function of \(I_2\) remains invariant under \(\ket{\Phi} \rightarrow U \otimes U \ket{\Phi}\). Due to parity conservation and \( \mathbb{Z}_2 \) symmetry in the TFIM, we have global spin-flip operations within each parity sector of the RDM. Additionally, a subregion may be symmetric under the permutation of sites which correspond to lattice symmetries such as rotations and reflections.
As the RDM acquires more symmetries, the \(I_2\) landscape becomes increasingly symmetric, simplifying the search for the optimal \(\ket{\Phi}\). 
While this symmetry in \(I_2\) does not uniquely determine the form of \(\ket{\Phi}\), empirically it often guides us to effective parametrizations for \(\ket{\Phi}\) that reduce the search space and enhance computational efficiency.

\paragraph{Three Adjacent Sites in 1d}

In the one-dimensional TFIM, the three-site RDM (indexed as sites 1, 2, and 3) is symmetric under the exchange of sites 1 and 3. 
This suggests a two-angle parametrization of \(\ket{\Phi}\):
\[
\begin{aligned}
    \theta_1 &= \theta_3 = -\theta_4 = -\theta_6 = x \\
    \theta_2 &= -\theta_5 = y 
\end{aligned}
\]
In 1d, this symmetry-constrained form yields optimal \(I_2\) values for $h\leq h_c$, and near-optimal values for $h>h_c$.

\paragraph{Four-Spin Plaquette in 2d/3d}

In 2d and 3d, we consider the four-spin plaquette RDM which exhibits rotational and reflectional symmetries. These symmetries suggest a unique angle parametrization:
\[
\begin{aligned}
\theta_1 &= \theta_2 = \theta_3 = \theta_4 = \theta, \\
\theta_5 &= \theta_6 = \theta_7 = \theta_8 = \frac{\pi}{2} - \theta
\end{aligned}
\]
In 2d ED with $L=5$, this one-angle parametrization yields optimal $I_2$ values for $h \leq 3.15$ and near-optimal values for larger $h$. In 3d ED with $L=3$, it provides optimal $I_2$ values within the range $3.9 \leq h \leq 5.8$ and remains near-optimal for $h>5.8$. For $h<3.9$, the one parameter $I_2$ is negative, while the full $I_2$ stays close to zero. This is an example where symmetry alone does not fully determine the structure of the optimal product state.  
Analysis on QMC data further confirms this one-angle parametrization as an effective ansatz. 


The symmetry-adapted approach should provide an efficient ansatz for \(I_2\) for large subregions, where a full parametrization would be computationally prohibitive. Symmetry-constrained forms of \(\ket{\Phi}\) can significantly reduce the number of parameters required, making it feasible to explore \(I_2\) in more complex systems. 

\section{Global multiparty entanglement}
In the previous section we analyzed the GME between 3 and 4 spins, but it turns out that we can exploit the tomography of small subregions to determine the most collective form of entanglement, that is the GME between \emph{all} spins in the lattice. More precisely, we shall evaluate the GMC (Eq.~\eqref{eq:gmc}) where the $N$-parties are the $N$ spins that form the lattice. For a general pure state of $N$ spins, one needs to obtain the minimum over all $2^{N-1}-1$ bipartitions $\gamma_i=(A_i|B_i)$. However, we can exploit the fact that the groundstate of the TFIM is uniform: by virtue of spatial symmetries, all sites are equivalent. The bipartition that has the smallest concurrence $\sqrt{2(1-\Tr\rho^2_{A_i})}$, i.e.\/ the least entanglement, will be when $A_i$ is a single site.
Indeed, adding more spins to $A_i$ will generate entanglement with more spins in $B_i$.
For instance, in the limit of a large subregion of adjacent sites $|A_i|\gg 1$, the purity decays exponentially: $\log_2\Tr\rho_{A_i}^2=-\alpha|\partial A_i|+\cdots$. In that case, the concurrence approaches its maximal value $C\to\sqrt2$, which certainly exceeds the maximal value for a single site subregion, $C=1$. We have numerically verified that lattices with $N\leq 20$ indeed always have their global minimum for single-site bipartitions. The single-site optimum allows us to write a remarkably simple closed-form result for the $N$-party GMC in the groundstate:
\begin{align} \label{eq:global-gmc}
    C_{\rm GME}(|\psi\rangle)\!= \sqrt{2(1-\Tr\rho_i^2)}= \sqrt{1-\langle\sigma_i^x\rangle^2-\langle\sigma_i^z\rangle^2}
\end{align}
where we have used $\langle\sigma_i^y\rangle=0$ in the TFIM by virtue of the complex-conjugation symmetry.
In addition, we have $\langle\sigma_i^z\rangle=0$ for all $h$ in a finite size system, so that $C_{\rm GME}$ solely depends on a simple physical observable: the single-site transverse magnetization $\langle\sigma_i^x\rangle$ (evaluated at any site, say $i=1$). We note that Eq.~\eqref{eq:global-gmc} leads to the expected singular scaling~\cite{wangEntanglement2024} as a function of the transverse field $h\to h_c$ discussed above.
It is important to remember that (\ref{eq:global-gmc}) holds for uniform states. For a general quantum state of $N$ spins, one has to optimize over an exponentially large number of bipartitions, and the result does not have a closed form.     

\begin{figure}
\centering
\includegraphics[width=1.06\columnwidth]{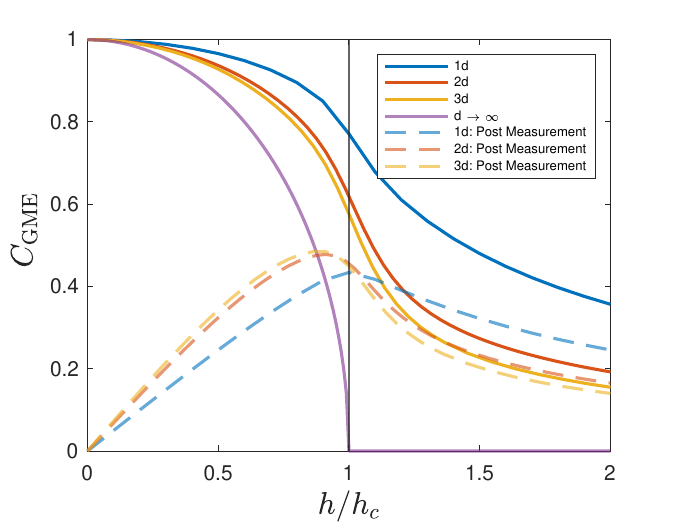}
\caption{\textbf{Global Genuine Multipartite Concurrence.} 
$N$-partite $C_{\rm GME}$ as a function of the field $h$ for all $N$ spins in the lattice, computed via ED for 1d $(L=25)$, 2d $(L=5)$ and 3d $(L=3)$ systems. The finite-size critical fields for these lattices are identified by the peak of $dS^{\rm vN}/dh$: $h_c^*$ = 1, 2.74, and 4.4 for 1d, 2d, and 3d lattices. 
Solid lines represent the GMC calculated for the entire system, while dashed lines depict the  GMC obtained after measuring $\sigma^z$ for one spin, and computing the GMC for the remaining  $N-1$  spins. The purple line represents the mean-field theory prediction in the limit $d\rightarrow \infty$.}
\label{fig:globalGMC}
\end{figure}

The result is plotted in solid lines for 1d, 2d, and 3d in Fig.~\ref{fig:globalGMC}, where we see that it is greater in 1d, just like the other entanglement measures considered in this work.
Further, we note that it interpolates between 1 at small fields, and zero at large $h$. Indeed, since we work with finite-size systems, the $h\to 0$ groundstate is the symmetric cat/GHZ state $|\psi\rangle=\tfrac{1}{\sqrt2}(|0\ldots 0\rangle+|1\dots 1\rangle)$, which indeed possesses $N$-party GME. However, in the presence of a small stray field, one would select one of the two ferromagnetic product states, and the GMC would vanish. The large GMC in the ferromagnet is thus fragile. We can obtain a more faithful measure of collective GME by checking the robustness of the GME to local measurements. The idea is to perform a projective measurement of $\sigma_i^z$ at site $i=1$, and then evaluate the $(N-1)$-party GMC among the unmeasured spins. The result is independent of both the choice of the measured spin (uniformity) and the measurement outcome (Ising $\mathbb Z_2$ symmetry). 
Post-measurement results in 1d, 2d, and 3d shown as dashed lines in Fig.~\ref{fig:globalGMC} exhibit a peak near the QCP, reflecting a trend similar to the local entanglement measures discussed in previous sections.


In the limit of large spatial dimension $d \rightarrow \infty$, mean-field theory predicts a critical transverse field $h_c=2 d$. For $h \geq h_c$, the system resides in the paramagnetic phase, characterized by $\left\langle\sigma_i^x\right\rangle=1$ and $\left\langle\sigma_i^z\right\rangle=0$, resulting in the absence of global entanglement, $C_{\mathrm{GME}}=0$. This behavior is consistent with the monotonic decrease of $C_{\mathrm{GME}}$ with increasing $d$, as shown in Fig. \ref{fig:globalGMC}. 
For $h<h_c$, the system transitions to the ferromagnetic phase, where $\left\langle\sigma_i^x\right\rangle=h / h_c$ and $\left\langle\sigma_i^z\right\rangle=\sqrt{1-\left(h / h_c\right)^2}$, corresponding to a symmetry-broken groundstate that also lacks global entanglement. 
However, for a finite $d$ and system size, the groundstate becomes a ferromagnetic cat state with $\langle\sigma_i^z\rangle=0$, resulting in a finite global entanglement $C_{\rm GMC} = \sqrt{1-\left(h / h_c\right)^2}$, which approaches $C_{\mathrm{GMC}} \rightarrow 1$ as $h \rightarrow 0$, as shown in Fig.~\ref{fig:globalGMC}.
This can be seen as the limit $d\rightarrow \infty$ for the pre-measured $C_{\rm GME}$.

Global entanglement in the TFIM has also been studies using the Meyer-Wallach measure~\cite{Montakhab2010}, which is a function of the average concurrence over all single-site subregions. 
Unlike the GMC, it is not a measure of GME but instead quantifies general entanglement, vanishing only for fully separable product states.
Despite this difference, for uniform states, the GMC has a simple relation to  the Meyer-Wallach measure, and their results in 1d, 2d, and 3d show similar behavior to our pre-measured global GMC, particularly the universal scaling near the QCP characterized by a divergence in the derivative, consistent with the universal singular scaling discussed in Sec.~\ref{section:singular_scaling}. Our study, however, extends to larger lattice sizes, and crucially includes post-measurement results, which yields additional insights into the robustness of global entanglement.



\section{Discussion}
\label{sec:secIV}

We began systematically charting the structure of bipartite and multipartite entanglement in the quantum-Ising models in 1d, 2d, and 3d via entanglement microscopy. 
For 4 sites or less, all forms of entanglement between nearby subregions (parties) peak near the quantum critical point.
When increasing the separation between the parties,  we find that 
both two-party and multi-party entanglement are extremely short-ranged. In particular, GME is shared among neighbouring subregions only. This demonstrates that, even at criticality, long-range correlations are predominantly classical. By increasing the spatial dimension, monogamy suppresses entanglement, with an even stronger impact in the multipartite case.
We further considered the most collective form of multipartite entanglement involving all spins, and showed that it reduces to the most local quantity---the entanglement between a single site and the rest of the system, which can be readily computed with entanglement microscopy. We have further shown that although the collective form is the largest in the ordered phase, a single measurement can drastically reduce it, and the post-measurement multipartite concurrence peaks near the transition.
These findings establish entanglement microscopy as a practical tool for diagnosing phases and criticality, and pave the way for applying the approach to more exotic models where entanglement may organize itself in fundamentally different ways.

Our work invites further investigation along various directions. First, it would be important to investigate larger subregions to understand more collective forms of entanglement.
In this respect, it would be  desirable to obtain precise RDMs of such subregions with QMC, where a careful handling of the symmetries would facilitate the simulations.
Second, it would be natural to employ entanglement microscopy to study the fate of entanglement as a function of temperature. For instance, upper bounds for the temperature at which GME can exist in the 1d Ising model were obtained for 3 adjacent spins~\cite{Wen2024}. One could extend these results to higher dimensions and to larger subregions. However, this comes with significant computational challenges. 
Our current practical limit for precision multipartite entanglement microscopy in 3d systems is \(L\approx 4\), with a required precision for the RDM matrix elements on the order of \(10^{-4}\) to \(10^{-5}\). Pushing beyond this limit to larger systems will require more advanced techniques.
One promising approach is to integrate tensor-network contraction into Monte Carlo sampling, as demonstrated for entanglement entropy in 2d systems~\cite{wangAnalog2024}. Extending such techniques to 3d systems would enable more precise entanglement studies in higher dimensions and larger subregions, offering deeper insights into the collective behavior of entanglement. 

The most fertile direction for future work would be to investigate other models in order to see what kind of distinct GME structures can arise compared to the TFIM. For instance, are there local Hamiltonians where multipartite entanglement is more collective in nature? 
Additionally, would the insights gained from quantum spin models extend to interacting fermion models, where particle statistics could introduce qualitatively different entanglement patterns? More broadly, a key question remains: how does microscopic entanglement shape and reflect the macroscopic properties of quantum many-body systems?

\section*{Acknowledgement}
MHS, TTW and ZYM acknowledge the support from the Research Grants Council (RGC) of Hong Kong (Project Nos. AoE/P-701/20, 17309822, HKU C7037-22GF, 17302223, 17301924), the ANR/RGC Joint Research Scheme sponsored by RGC of Hong Kong and French National Research Agency (Project No. A\_HKU703/22). 
W.W.-K.\/ and L.L.\/ are supported by a grant from the Fondation Courtois, a Chair of the Institut Courtois, a Discovery Grant from NSERC, and a Canada Research Chair.
We thank HPC2021 system under the Information Technology Services and the Blackbody HPC system at the Department of Physics, University of Hong Kong, as well as the Beijng PARATERA Tech CO.,Ltd. (URL: https://cloud.paratera.com) for providing HPC resources that have contributed to the research results reported within this paper.

\bibliographystyle{longapsrev4-2}
\bibliography{bibtext}

\clearpage
\onecolumngrid

\section{Appendix}

\subsection{RDMs from Entanglement Microscopy} \label{Appendix:RDM_matrox}

This section presents a selection of RDMs at the QCP in 2d and 3d, obtained from both ED and QMC. Additional RDMs from QMC can be found on the \href{https://github.com/songmengh/Ising_RDMs/tree/main}{GitHub repository}. Note that the RDMs on GitHub follow the original convention, \( H = -J \sum_{\langle i, j \rangle} \sigma_i^z \sigma_j^z - h \sum_i \sigma_i^x \), while all RDMs shown in this section have been transformed by a local unitary—specifically, a rotation by \(\pi/2\) around the \(Y\)-axis—to effectively map to the Ising Hamiltonian with \( \sigma^x \leftrightarrow \sigma^z \) exchanged: \( H = -J \sum_{\langle i, j \rangle} \sigma_i^x \sigma_j^x - h \sum_i \sigma_i^z \).
This transformation results in sparser RDMs, making the comparison between ED and QMC more straightforward.

Below, we provide RDMs of three adjacent spins (configured as 124, see Fig.~\ref{fig:fig1} for site indices) at the QCP in 2d and 3d, from both ED and QMC.

\subsubsection{2d}

For the 2d case, we present the RDMs:

ED, lattice size $6\times5$, $h=3$
\[
\begin{bmatrix}
    0.8323 & 0       & 0       & 0.0413 & 0       & 0.0478 & 0.0979 & 0       \\
    0       & 0.0539 & 0.0364 & 0       & 0.0399 & 0       & 0       & 0.0074 \\
    0       & 0.0364 & 0.0459 & 0       & 0.0366 & 0       & 0       & 0.0059 \\
    0.0413 & 0       & 0       & 0.0041 & 0       & 0.0039 & 0.0060 & 0       \\
    0       & 0.0399 & 0.0366 & 0       & 0.0453 & 0       & 0       & 0.0061 \\
    0.0478 & 0       & 0       & 0.0039 & 0       & 0.0047 & 0.0067 & 0       \\
    0.0979 & 0       & 0       & 0.0060 & 0       & 0.0067 & 0.0127 & 0       \\
    0       & 0.0074 & 0.0059 & 0       & 0.0061 & 0       & 0       & 0.0011 \\
\end{bmatrix}
\]

QMC, lattice size $24 \times 24$, $h=3.044$
\[
\begin{bmatrix}
    0.8732 & 0       & 0       & 0.0409 & 0       & 0.0904 & 0.0904 & 0       \\
    0       & 0.0363 & 0.0245 & 0       & 0.0232 & 0       & 0       & 0.0057 \\
    0       & 0.0245 & 0.0362 & 0       & 0.0232 & 0       & 0       & 0.0057 \\
    0.0409 & 0       & 0       & 0.0032 & 0       & 0.0049 & 0.0049 & 0       \\
    0       & 0.0232 & 0.0232 & 0       & 0.0293 & 0       & 0       & 0.0045 \\
    0.0904 & 0       & 0       & 0.0049 & 0       & 0.0102 & 0.0099 & 0       \\
    0.0904 & 0       & 0       & 0.0049 & 0       & 0.0099 & 0.0103 & 0       \\
    0       & 0.0057 & 0.0057 & 0       & 0.0045 & 0       & 0       & 0.0014 \\
\end{bmatrix}
\]

\subsubsection{3d}

For 3d, we present the RDMs:

ED, lattice size $3 \times 3 \times 3$, $h=5.2$
\[
\begin{bmatrix}
    0.9130 & 0       & 0       & 0.0641 & 0       & 0.0279 & 0.0641 & 0       \\
    0       & 0.0262 & 0.0192 & 0       & 0.0191 & 0       & 0       & 0.0030 \\
    0       & 0.0192 & 0.0227 & 0       & 0.0192 & 0       & 0       & 0.0026 \\
    0.0641 & 0       & 0       & 0.0050 & 0       & 0.0024 & 0.0048 & 0       \\
    0       & 0.0191 & 0.0192 & 0       & 0.0262 & 0       & 0       & 0.0030 \\
    0.0279 & 0       & 0       & 0.0024 & 0       & 0.0015 & 0.0024 & 0       \\
    0.0641 & 0       & 0       & 0.0048 & 0       & 0.0024 & 0.0050 & 0       \\
    0       & 0.0030 & 0.0026 & 0       & 0.0030 & 0       & 0       & 0.0004 \\
\end{bmatrix}
\]

QMC, lattice size $8 \times 8 \times 8$, $h=5.2$
\[
\begin{bmatrix}
    0.9427 & 0       & 0       & 0.0150 & 0       & 0.0530 & 0.0530 & 0       \\
    0       & 0.0176 & 0.0082 & 0       & 0.0084 & 0       & 0       & 0.0014 \\
    0       & 0.0082 & 0.0176 & 0       & 0.0084 & 0       & 0       & 0.0014 \\
    0.0150 & 0       & 0       & 0.0006 & 0       & 0.0010 & 0.0010 & 0       \\
    0       & 0.0084 & 0.0084 & 0       & 0.0150 & 0       & 0       & 0.0010 \\
    0.0530 & 0       & 0       & 0.0010 & 0       & 0.0032 & 0.0031 & 0       \\
    0.0530 & 0       & 0       & 0.0010 & 0       & 0.0031 & 0.0032 & 0       \\
    0       & 0.0014 & 0.0014 & 0       & 0.0010 & 0       & 0       & 0.0002 \\
\end{bmatrix}
\]


\subsection{Proof that $I_2$ is strictly stronger than $W_1$ criterion} \label{Appendix:I2 stronger than W1}
In this section, we will show that for three spins, the \( W_{1} \) criterion, obtained by optimizing over either local unitary or local filtering matrices, can be obtained by restricting the parameter space of the \( I_{2} \) criterion. This will demonstrate that \( I_{2} \) is strictly stronger than \( W_{1} \).

First, define the objective function
\begin{equation*}
    f(\rho, \vert \Phi \rangle) := \sqrt{\langle\Phi| \rho^{\otimes 2} \Pi|\Phi\rangle} - \sum_i \sqrt{\langle\Phi| \mathcal{P}_i^{\dagger} \rho^{\otimes 2} \mathcal{P}_i|\Phi\rangle}
\end{equation*}

Notice that $f$ scales linearly with $\rho$ and $\vert \Phi \rangle$: for $c_{\rho}, c_{\Phi} > 0$, we have
\begin{equation*}
    f(c_{\rho} \rho, c_{\Phi} \vert \Phi \rangle) = c_{\rho} c_{\Phi} f( \rho,  \vert \Phi \rangle)
\end{equation*}

Now, consider local filtering matrices \( F = F_1 \otimes F_2 \otimes \dots \otimes F_n \) on $\mathcal{H}_1 \otimes \cdots \otimes \mathcal{H}_n$, where \( F_i \) is an arbitrary complex invertible matrix. Since permutation operators \(\Pi\) and \(\mathcal{P}_i\) only permute identical copies, we have \([F^{\otimes 2}, \Pi] = [F^{\otimes 2}, \mathcal{P}_i] = 0\) for all \( i \). This leads to the filtering property:
\begin{equation*}
    f(F^\dagger \rho F, \vert \Phi \rangle) = f(\rho, F^{\otimes 2} \vert \Phi \rangle)
\end{equation*}

This equation also holds if we restrict the filtering matrix to unitary matrices \( F_i = U_i \). We can thus rewrite the maximization in terms of local unitary transformations:
\begin{align*}
    I_2[\rho] &= \max_{\ket{\Phi}} f(\rho, \ket{\Phi}) \\
              &= \max_{U_A, U_B} f(\rho, U_A \otimes U_B \ket{\Phi^0}) \\
              &\geq \max_{U} f(\rho, U^{\otimes 2} \ket{\Phi^0}) \\
              &= \max_{U} f(U^\dagger \rho U, \ket{\Phi^0})
\end{align*}
where \( \ket{\Phi^0} \) is an arbitrary fixed product state. By restricting to \( U_A = U_B = U \), we obtain an inequality. For three spins, if we choose \( \ket{\Phi^0} = \ket{000111} \), we recover \( W_1 \):
\begin{align*}
    W_1[\rho] &= \max_{U} f(U^\dagger \rho U, \ket{000111}) \\
              &= \max_{U} \Bigg\{ 
 \left|\left\langle 000\right|U^{\dagger}\rho U\left|111\right\rangle \right| - \sqrt{\left\langle 100\right|U^{\dagger}\rho U\left|100\right\rangle \left\langle 011\right|U^{\dagger}\rho U\left|011\right\rangle} \\
&\quad - \sqrt{\left\langle 010\right|U^{\dagger}\rho U\left|010\right\rangle \left\langle 101\right|U^{\dagger}\rho U\left|101\right\rangle} \\
&\quad - \sqrt{\left\langle 001\right|U^{\dagger}\rho U\left|001\right\rangle \left\langle 110\right|U^{\dagger}\rho U\left|110\right\rangle}
\Bigg\}
\end{align*}

To obtain the canonical form in Ref~\cite{Guhne_2010}, we identify the indices \(\{000,001,010,011,100,101,110,111\}\) with \(\{1,2,3,4,5,6,7,8\}\). This shows that for any state \( \rho \), when both criteria are optimized over local unitary matrices, \( I_{2}[\rho] \geq W_{1}[\rho] \).

We can extend this proof to the case of filtering, showing \( I_{2F}[\rho] \geq W_{1F}[\rho] \):
\begin{align*}
    I_{2F}[\rho] &= \max_{\ket{\Phi}, F} f(F^\dagger \rho F, \ket{\Phi}) \\
                 &= \max_{U_A, U_B, F} f(F^\dagger \rho F, U_A \otimes U_B \ket{\Phi^0}) \\
                 &\geq \max_{U, F} f(\rho, U^{\otimes 2} \ket{\Phi^0}) \\
                 &= \max_{U, F} f(U^\dagger F^\dagger \rho F U, \ket{\Phi^0}) \\
                 &= \max_{F} f(F^\dagger \rho F, \ket{\Phi^0})
\end{align*}
where in the last step, we absorbed the unitary matrix into the more general filtering matrix. Choosing \( \ket{\Phi^0} = \ket{000111} \) again yields:
\begin{equation*}
    W_{1F}[\rho] = \max_{F} f(F^\dagger \rho F, \ket{000111})
\end{equation*}
proving that \( I_{2F}[\rho] \geq W_{1F}[\rho] \).

Finally, we show that \( I_{2F}[\rho] \) and \( I_2[\rho] \) always have the same sign for a quantum state \( \rho \). Since
\begin{align*}
    I_{2F}[\rho] &= \max_{\ket{\Phi}, F} f(F^\dagger \rho F, \ket{\Phi}) \\
                 &= \max_{\ket{\Phi}, F} f(\rho, F^{\otimes 2} \ket{\Phi}),
\end{align*}
we see that if \( f(\rho, F^{\otimes 2} \ket{\Phi}) > 0 \) for some \( \ket{\Phi}, F \), then normalizing \( F^{\otimes 2} \ket{\Phi} = \sqrt{\langle \Phi | (F^\dagger F)^{\otimes 2} | \Phi \rangle} \ket{\Phi'} \) gives:
\begin{equation*}
    f(\rho, F^{\otimes 2} \ket{\Phi}) = \sqrt{\langle \Phi | (F^\dagger F)^{\otimes 2} | \Phi \rangle} f(\rho, \ket{\Phi'}),
\end{equation*}
implying that there exists \( \ket{\Phi'} \) such that \( f(\rho, \ket{\Phi'}) > 0 \). Thus, filtering does not make the criterion stronger.

\begin{figure}[htp!]
    \centering
    \includegraphics[width=0.6\columnwidth]{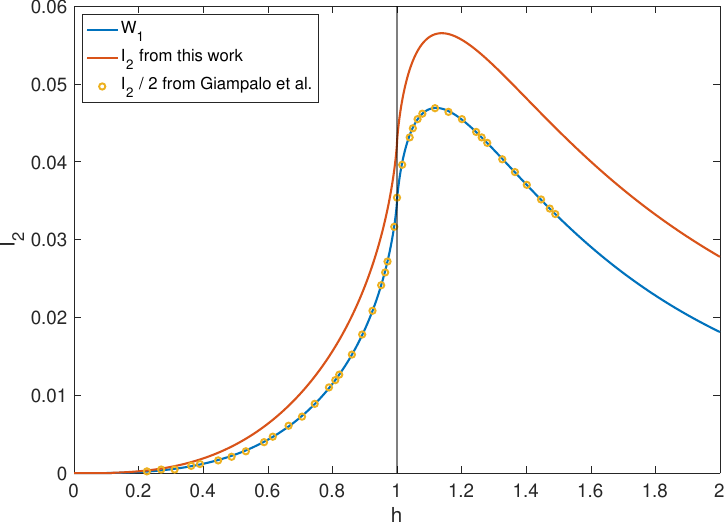}
   \caption{Comparison of the GME measures \( I_2 \) and \( W_1 \) as functions of the transverse field \( h \) for the 1d Ising model. The solid blue line represents the \( W_1 \) criterion computed in our previous work~\cite{wangEntanglement2024}, while the red line shows our optimized \( I_2 \) criterion. The orange points correspond to the values of \( I_2 / 2 \) extracted from Ref.~\cite{Giampaolo2013} (Fig. 1, \( \gamma = 1 \) curve), which fall directly on the \( W_1 \) curve. This alignment indicates that Ref.~\cite{Giampaolo2013} performed their optimization only over local unitary transformations, effectively implementing the \( W_1 \) criterion and multiplying it by 2. In contrast, our approach optimizes \( I_2 \) over all product states \( | \Phi \rangle \), making \( I_2 \) a strictly stronger criterion than \( W_1 \) for detecting GME, as demonstrated by \( I_2 \geq W_1 \) for all \( h \). At the QCP, \( I_2 = 0.0425 \) and \( W_1 = 0.0350 \).}
    \label{fig:I2_W1_comparison}
\end{figure}

As shown in Fig.~\ref{fig:I2_W1_comparison}, this comparison clearly demonstrates that the $I_2$ criterion is strictly stronger than the $W_1$ criterion for the 1D Ising model. Notably, as discussed in Section~\ref{section:singular_scaling} of the main text, the figure also indicates that the results of Ref.~\cite{Giampaolo2013} coincide with the $W_1$ curve, confirming that their optimization was effectively limited to local unitary transformations (i.e., the $W_1$ criterion), rather than the full space of product states required by $I_2$. This highlights the improved detection power of $I_2$ for GME.

\subsection{Proof that $2I_2$ lower bounds $C_{\rm GME}$ for an arbitrary product state $\ket{\Phi}$}\label{Appendix:lower_bound_proof}

In this section we amend the proof given in Ref.~\cite{Ma2011} by showing that $2 I_2[\rho, \ket{\Phi}] \leq C_{\rm GME}(\rho)$ holds for an arbitrary product state $\ket{\Phi}$. We will first consider the case of three spins. In Ref.~\cite{Ma2011}, the following inequality was proved first for pure state and then for mixed states using convexity:
$$C_{\rm GME}(\rho) \geqslant 2 I[\rho,|000111\rangle]$$
under an optimization over local unitary transformations, the right-hand side reduces to the $W_1$ criteria, see \ref{Appendix:I2 stronger than W1}, while the left-hand side stays invariant. Therefore, the lower bound was only proved for $W_1$:  $C_{\rm GME}(\rho) \geqslant 2 W_1(\rho)$ which is weaker than $I_2$. To prove that it holds for $I_2$, we need to prove that
$$C_{\rm GME}(\rho) \geqslant 2 I[\rho,|000xyz\rangle]$$
where $\ket{x},\ket{y},\ket{z}$ are arbitrary states. We now show that this is true for pure states. 

For pure states, $I_2$ simplifies to
$$I_2[|\psi\rangle\langle\psi|,|\Phi\rangle]=\left|\left\langle\Phi_1 \mid \psi\right\rangle\left\langle\Phi_2 \mid \psi\right\rangle\right|-\sum_{\gamma_i}\left|\left\langle\tilde{\Phi}_{i 1} \mid \psi\right\rangle\left\langle\tilde{\Phi}_{i 2} \mid \psi\right\rangle\right|$$
where $\Ket{\Phi} = \Ket{\Phi_1 \Phi_2}$, for each bipartition $\gamma_i$ we define product states
$\mathcal{P}_i \Ket{\Phi} = \Pi_{A_i} \otimes \mathbbm{1}^{B_i} \Ket{\Phi_1 \Phi_2} = \Ket{\tilde{\Phi}_{i, 1} \tilde{\Phi}_{i, 2}}$. 
Consider $\Ket{\Phi} = \Ket{000xyz}$, we have
$$
\begin{aligned}
\left|\Phi_1\right\rangle & =|000\rangle, & & \left|\Phi_2\right\rangle=|x y z\rangle \\
\left|\tilde{\Phi}_{1,1}\right\rangle & =|x 00\rangle, & & \left|\tilde{\Phi}_{1,2}\right\rangle=|0 y z\rangle \\
\left|\tilde{\Phi}_{2,1}\right\rangle & =|0 y 0\rangle, & & \left|\tilde{\Phi}_{1,2}\right\rangle=|x 0 z\rangle \\
\left|\tilde{\Phi}_{3,1}\right\rangle & =|00 z\rangle, & & \left|\tilde{\Phi}_{1,2}\right\rangle=|x y 0\rangle
\end{aligned}
$$

observe that both the original and permuted product states can be expanded as
$$
\begin{aligned}
\left|\Phi_1\right\rangle\left|\Phi_2\right\rangle & =\sum_{i j k} x_i y_j z_k|000 i j k\rangle \\
\left|\tilde{\Phi}_{k 1}\right\rangle\left|\tilde{\Phi}_{k 2}\right\rangle & =\sum_{i j k} x_i y_j z_k\left(\Pi_{A_k} \otimes I^{B_k}|000 i j k\rangle\right)
\end{aligned}
$$

where $|x_1|^2+|x_2|^2=|y_1|^2+|y_2|^2=|z_1|^2+|z_2|^2=1$. Using this form, we can write $I_2$ as

$$\begin{aligned}
I_2[|\psi\rangle\langle\psi|,|000 x y z\rangle] & =\sum_{i j k}\left\{\left|x_i y_j z_k a_{000} a_{i j k}\right|-\left|x_i y_j z_k a_{i 00} a_{0 j k}\right|-\left|x_i y_j z_k a_{0 j 0} a_{0 j 0}\right|-\left|x_i y_j z_k a_{00 k} a_{i j 0}\right|\right\} \\
& =\sum_{i j k}\left|x_i y_j z_k\right|\left\{\left|a_{000} a_{i j k}\right|-\left|a_{i 00} a_{0 j k}\right|-\left|a_{0 j 0} a_{0 j 0}\right|-\left|a_{00 k} a_{i j 0}\right|\right\} \\
& =\sum_{i j k}\left|x_i y_j z_k\right| f_{i j k} \\
f_{i j k}&=\left|a_{000} a_{i j k}\right|-\left|a_{i 00} a_{0 j k}\right|-\left|a_{0 j 0} a_{0 j 0}\right|-\left|a_{00 k} a_{i j 0}\right|
\end{aligned}
$$

note that $f_{i j k} \leq 0$ for $(i,j,k)\neq (1,1,1)$. If $f_{111}\leq0$, then $I_2<0$. If $f_{111}>0$, the maximum is simply $f_{111}$,  achieved by $\left|x_1\right|=\left|y_1\right|=\left|z_1\right|=1$. This means that 

$$
\max \left\{0, \max _{|x y z\rangle} I_2[|\psi\rangle\langle\psi|,|000 x y z\rangle]\right\}=\max \left\{0, I_2[|\psi\rangle\langle\psi|,|000111\rangle]\right\}
$$

Therefore, there is no loss of generality in choosing $\Ket{000111}$ as the product state. Following Ref.~\cite{Ma2011}, we extend the inequality to 

$$
C_{G M E}(|\psi\rangle) \geqslant \max \{0,2 I[|\psi\rangle\langle\psi|,|000111\rangle]\}=\max \left\{0, I_2[|\psi\rangle\langle\psi|,|000 x y z\rangle]\right\}
$$

Since the left-hand side is invariant under local unitary transformation, we have 

$$\begin{aligned}
C_{G M E}(|\psi\rangle) & \geq \max \left\{0, \max _U I_2\left[U|\psi\rangle\langle\psi| U^{\dagger},|000 x y z\rangle\right]\right\} \\
& =\max \left\{0, \max _U I_2[|\psi\rangle\langle\psi|, U \otimes U|000 x y z\rangle]\right\} \\
& =\max \left\{0, \max _{|\Phi\rangle} I_2[|\psi\rangle\langle\psi|,|\Phi\rangle]\right\}
\end{aligned}$$
where we used the fact that any product state $\Ket{\Phi}$ can always be written as 
$$
|\Phi\rangle=U|000\rangle \otimes U|x y z\rangle=\left(U_1 \otimes U_2 \otimes U_3|000\rangle\right) \otimes\left(U_1 \otimes U_2 \otimes U_3|x y z\rangle\right)
$$
for some $U_1, U_2, U_3$ and $|x y z\rangle$. Using the convexity of $I_2$, we can generalize this lower bound to mixed states:
$$
\begin{aligned} C_{G M E}(\rho) & =\inf _{\left\{p_i,\left|\psi_i\right\rangle\right\}} \sum_i p_i C_{G M E}\left(\left|\psi_i\right\rangle\right) \\ & \geq 2 \inf _{\left\{p_i,\left|\psi_i\right\rangle\right\}} \sum_i p_i \max \left\{0, I_2\left(\left|\psi_i\right\rangle\left\langle\psi_i\right|\right)\right\} \\ & \geq \max \left\{0,2 I_2\left(\sum_i p_i\left|\psi_i\right\rangle\left\langle\psi_i\right|\right)\right\} \\ & \geq \max\{0, 2 I_2(\rho) \}
\end{aligned}
$$

The proof generalizes to \( N \) spins, where each subsystem is a qubit (\( d = 2 \)). For higher-dimensional subsystems (\( d > 2 \)), the product state \(\ket{\Phi}\) should be chosen as \(\ket{\Phi} = \Ket{00\ldots 0x^1x^2\ldots x^k}\), where \(\ket{x^k} = \sum_i x^k_i \ket{i}\). However, any \(\ket{x^k}\) can be reduced via a local unitary transformation to the subspace spanned by \(\{\ket{0}, \ket{1}\}\). 
This allows us to restrict \(\ket{x^k}\) to the qubit subspace and extend the lower bound to arbitrary dimensions by following the same steps as in the qubit case.


\subsection{Singular Scaling vs field} \label{Appendix:Singular Scaling}
As discussed in the main text, the derivative of a generic entanglement measure exhibits divergent behavior near the QCP, even for measures that rely on optimization, such as $I_2$. Specifically, the derivative of any entanglement measure $\mathcal{M}$ scales as $\frac{d \mathcal{M}}{d h} = \alpha |h - h_c|^{\Delta_{\varepsilon} \nu - 1}$ as $h \rightarrow h_c$. This appendix presents specific instances of this general scaling for $I_2$, the order parameter $\langle \sigma^x \rangle$ (related to the scaling of entanglement entropy), and Log Negativity.

\begin{figure}[htp!]
    \centering
    \includegraphics[width=0.8\columnwidth]{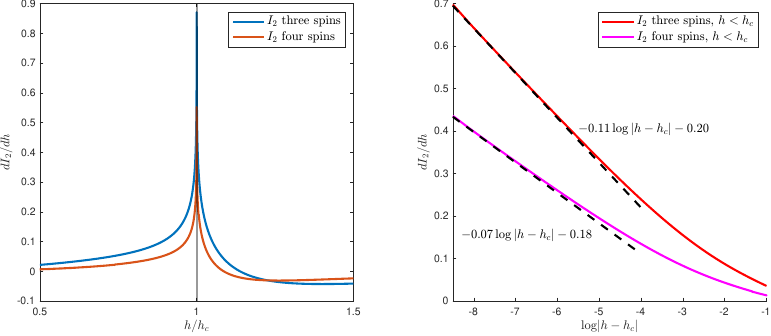}
    \caption{\textbf{Derivative of $I_2$ vs field $h$ for 1d TFIM}. The results are shown for three spins and four spins, respectively. The curves were fitted for $h < h_c$ in the range $\log_2 |h - h_c| \in [-8, -7]$.}
    \label{fig:I2_1d_logscale}
\end{figure}

\begin{figure}[htp!]
\centering
\includegraphics[width=0.5\columnwidth]{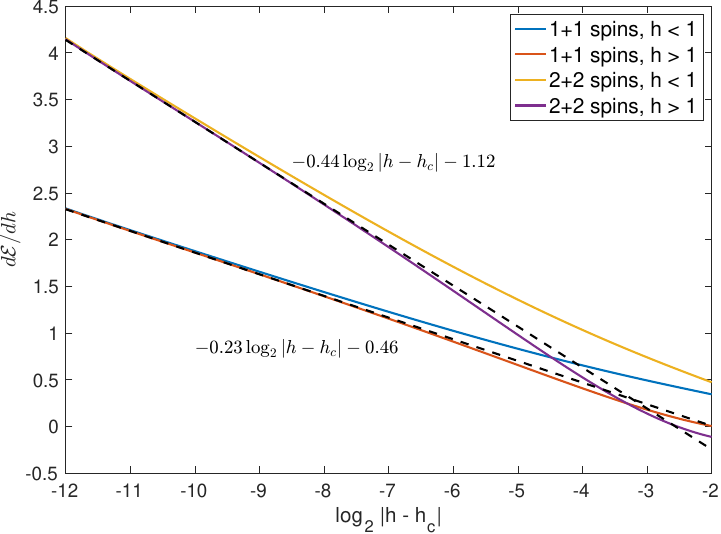}
\caption{The derivative of the Log Negativity, $\frac{d \mathcal{E}}{dh}$, as a function of $\log_2 |h - h_c|$ for two configurations of spins: $1+1$ and $2+2$. Solid lines correspond to the numerical data, while dashed lines represent linear fits for $h > 1$ in the range $\log_2 |h - h_c| \in [-12, -9]$.}
\label{fig:LN_log_scale}
\end{figure}

For 1d, the logarithmic divergence of $I_2$ near the QCP, as shown in Fig.~\ref{fig:I2_1d_logscale}, aligns with the expected logarithmic scaling at criticality. Similarly, the derivative of Log Negativity $\frac{d \mathcal{E}}{d h}$ follows a comparable scaling, illustrated in Fig.~\ref{fig:LN_log_scale}, with results for both two spins and two pairs of adjacent spins. 

\begin{figure}[htp!]
\centering
\includegraphics[width=0.5\columnwidth]{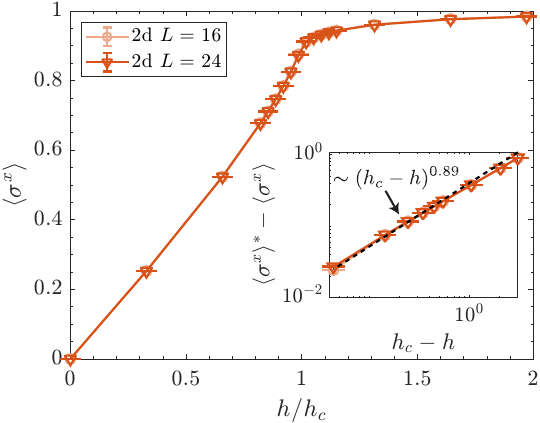}
\caption{\textbf{$\langle \sigma_x \rangle$ vs field $h/h_c$ for 2d TFIM.} The results are shown for lattice sizes $L=16$ and $L=24$ using QMC simulations. The scaling $\langle \sigma_x \rangle \sim |h - h_c|^{0.89}$ (up to an additive constant $\langle \sigma_x \rangle^*$ which is the $\langle \sigma_x \rangle$ value at the critical point.) is observed, supporting the entanglement entropy scaling in 2d: $\frac{d S^{\mathrm{vN}}}{d h} \sim\left|h-h_c\right|^{-0.11}$.}
\label{fig:2d_sigma_x}
\end{figure}

The order parameter $\langle \sigma^x \rangle$, which determines the one-site entanglement entropy, is shown here specifically for the 2d TFIM. It exhibits the expected scaling behavior $\langle \sigma^x \rangle \sim |h - h_c|^{0.89}$ (up to an additive constant $\langle \sigma_x \rangle^*$), as illustrated in Fig.~\ref{fig:2d_sigma_x}. This result supports the entanglement entropy scaling form $\frac{d S^{\mathrm{vN}}}{d h} \sim |h - h_c|^{-0.11}$ derived in the main text.

\begin{figure}[htp!]
    \centering
    \includegraphics[width=1\columnwidth]{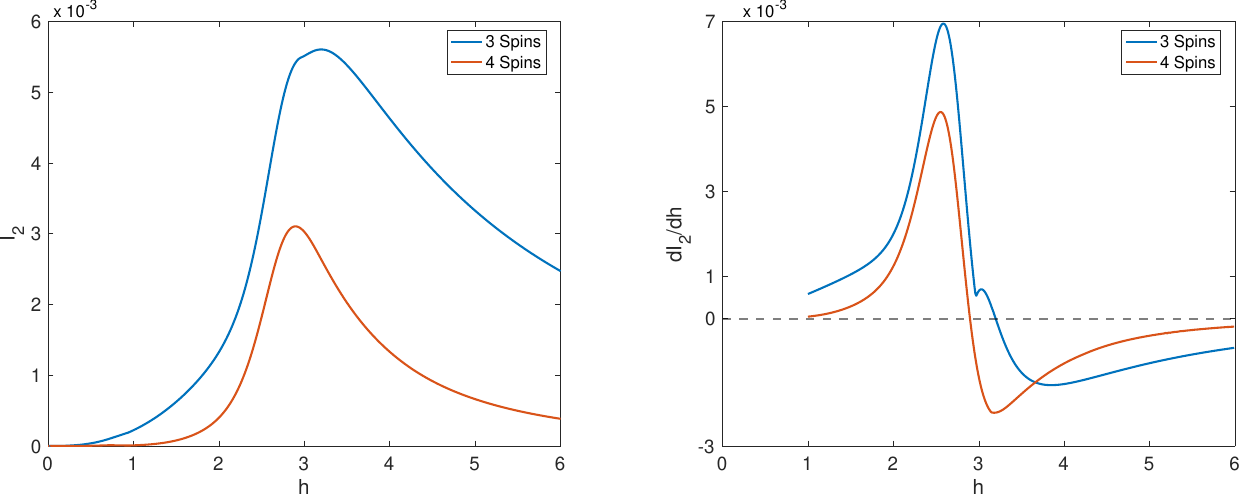}
  \caption{(Left) $I_2$ as a function of the transverse field $h$ for a 2d lattice of size $5 \times 5$. The results are shown for three spins and four spins, respectively. (Right) The derivative $dI_2/dh$ as a function of $h$. The peak of the derivative occurs at $h = 2.57$.}
    \label{fig:I2_vs_h_2d_TFIM}
\end{figure}

\begin{figure}[htp!]
    \centering
     \includegraphics[width=1\columnwidth]{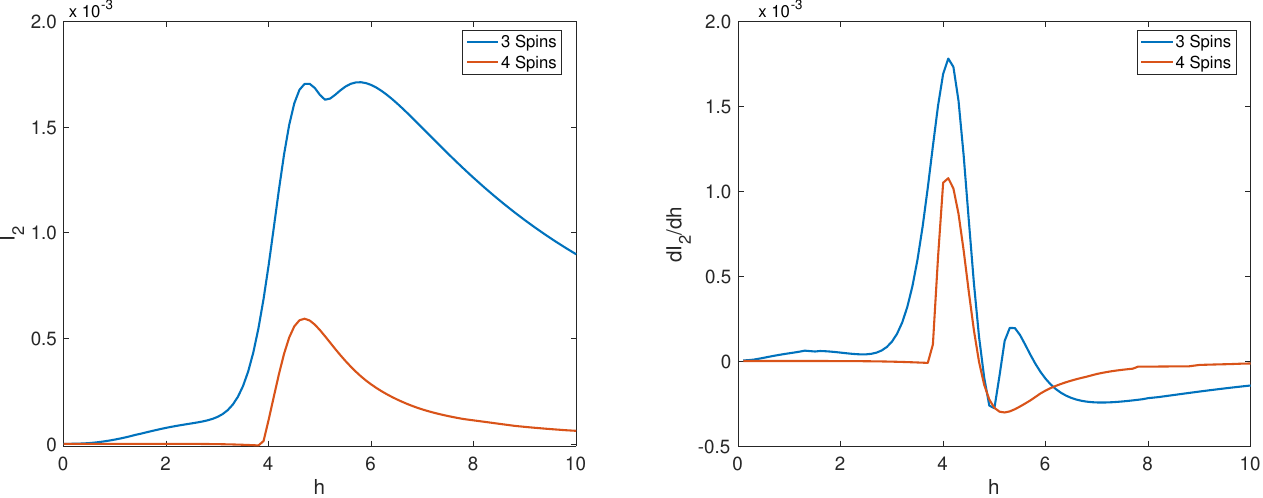}
   \caption{(Left) $I_2$ as a function of the transverse field $h$ for a 3d lattice of size $3 \times 3 \times 3$. The results are shown for three spins and four spins, respectively. (Right) The derivative $dI_2/dh$ as a function of $h$. The peak of the derivative occurs at $h = 4.1$.}
    \label{fig:I2_vs_h_3d_TFIM}
\end{figure}

For higher dimensions, specifically 2d and 3d, $I_2$ is expected to exhibit analogous critical behavior. In 2d, $\frac{d I_2}{d h}$ should follow power-law scaling near the QCP, while in 3d, a logarithmic divergence similar to that seen in 1d is anticipated. Figs.\ref{fig:I2_vs_h_2d_TFIM} and\ref{fig:I2_vs_h_3d_TFIM} display ED results for 2d and 3d, showing $I_2$ and its derivative as functions of $h$. In both cases, the peak of $\frac{d I_2}{d h}$ is observed close to the QCP, consistent with findings from the main text. We note that finite lattice sizes cause the apparent QCP to deviate slightly from the thermodynamic limit, though the proximity of these peaks to the QCP still suggests critical behavior. It is also acknowledged that the lattices here are too small to reveal true divergent behavior. While QMC methods can handle larger lattices, obtaining enough data points with sufficiently small error bars to capture the divergence accurately remains challenging.

\begin{figure}[htp!]
    \centering
     \includegraphics[width=0.5\columnwidth]{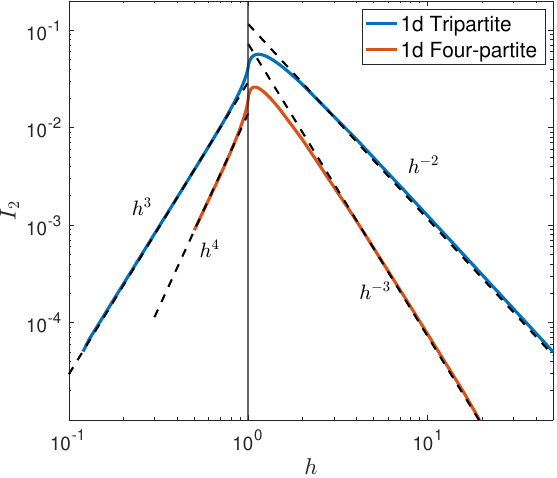}
   \caption{Log-log plot of $I_2$ as a function of $h$ for the 1d TFIM. 
    The solid lines represent $I_2$ for tripartite (3 sites) and four-partite (4 sites) entanglement on a 1d infinite lattice.
    The dashed lines indicate the expected scaling behavior at small and large fields based on perturbation analysis:
    $h^3$ and $h^{-2}$ for tripartite entanglement, and $h^4$ and $h^{-3}$ for four-partite entanglement. }
    \label{fig:I2_vs_h_power}
\end{figure}

We now discuss the asymptotic scaling of $I_2$ vs $h$ for small and large fields. For the 1d TFIM, the log-log plot of  $I_2$  as a function of $h$ (Fig.~\ref{fig:I2_vs_h_power}) demonstrates the expected scaling behavior at both small and large fields, consistent with the perturbative analysis presented in the main text. Specifically, for tripartite entanglement,  $I_2$  scales as  $h^3$  at small fields and  $h^{-2}$  at large fields, while for four-partite entanglement, the exponents are  $h^4$  and  $h^{-3}$ . These exponents are clearly observed in the data, as highlighted by the dashed lines. For 2d and 3d, we expect the same scaling exponents. 

\end{document}